\def\lesssim{\mathrel{\hbox{\rlap{\hbox{\lower4pt\hbox{$\sim$}}}\hbox{$<$}}}}

\def\gtrsim{\mathrel{\hbox{\rlap{\hbox{\lower4pt\hbox{$\sim$}}}\hbox{$>$}}}}

\def\msun{$M_{\odot}$}

\def\ll_lsun{log$({L/\rm L_{\odot}})$~}

\def\masa_msun{$M/ \rm M_{\odot}$~}

\def\m_mstar{$M/M_{*}$~}

\documentclass{aa}
\usepackage{graphicx}
        
\begin{document}

\title{The rate of period change in pulsating DB white dwarf stars} 

\author{A. H.    C\'orsico$^{1,2}$\thanks{Member    of    the    
        Carrera    del
        Investigador   Cient\'{\i}fico   y   Tecnol\'ogico,   CONICET,
        Argentina.},   L.     G.
        Althaus$^{1,2,3}$\thanks{Member    of    the    Carrera    del
        Investigador   Cient\'{\i}fico   y   Tecnol\'ogico,   CONICET,
        Argentina.}}

\offprints{A. H. C\'orsico}

\institute{$^1$Facultad de Ciencias Astron\'omicas y  Geof\'{\i}sicas, 
	   Universidad Nacional  de La  Plata, Paseo del  Bosque, s/n,
	   (1900) La    Plata,     Argentina.\\    $^2$Instituto    de
	   Astrof\'{\i}sica La  Plata, IALP, CONICET\\ $^3$Departament
	   de  F\'\i  sica   Aplicada,  Universitat  Polit\`ecnica  de
	   Catalunya,   Av.  del   Canal  Ol\'\i   mpic,   s/n,  08860
	   Castelldefels, Spain\\
\email{acorsico@fcaglp.unlp.edu.ar,      leandro@fa.upc.es}}

\date{Received; accepted}

\abstract{{In this work, we present the theoretically expected rates of 
pulsation period change for V777 Her (DBV) variable stars. To this end
we employ new evolutionary models representative of pulsating DB white
dwarf stars computed in a self-consistent way with the predictions of
time-dependent  element  diffusion.   At   the  hot  edge  of  the  DB
instability strip, the envelopes of  the models are characterized by a
diffusion-induced  double-layered chemical  structure. We  compute the
numerical values of rates of period change by solving the equations of
linear,  adiabatic, nonradial  stellar oscillations.   We  examine the
effects of varying  the stellar mass, the mass  of the helium envelope
and the neutrino  emission on the expected period  changes. We present
extensive  tabulations  of  our  results  which could  be  useful  for
comparison  with future  detections of  the rate  of period  change in
pulsating DB white dwarfs.}
 }

\authorrunning{C\'orsico \& Althaus}

\titlerunning{The rate of period change in DBV white dwarf stars}

\maketitle


\section{Introduction}

V777 Her  (or DBV) stars  are $g$(gravity)-mode variable  white dwarfs
with   helium-rich  atmospheres   (DB)   and  intermediate   effective
temperature  ($T_{\rm eff}  \sim 25000$  K), the  pulsating  nature of
which was predicted  two decades ago on theoretical  grounds by Winget
et al.  (1982a) and  shortly after confirmed observationally by Winget
et al.  (1982b). Since  then, considerable effort  has been  devoted to
studying these  stars. In particular,  the multiperiodic star  GD 358,
the most  extensively studied  member of the  DBV class, has  been the
subject  of  numerous  investigations  devoted  to  disentangling  its
internal structure  and evolution, initially  by means of  ``hand on''
asteroseismological procedures  (Bradley \& Winget 1994)  and later by
employing  objective fitting  techniques  (see, e.g,  Metcalfe et  al.
2000; 2001;  2002).  In  particular, Metcalfe et  al.  (2001)  --- see
also  Metcalfe  et  al.  (2002)  ---  have  recently  applied  genetic
algorithm-based    procedures   to    place    constraints   on    the
$^{12}$C$(\alpha,\gamma)^{16}$O reaction rate  from inferences for the
abundance of central oxygen in GD 358.

\begin{table*}
\centerline{TABLE 1}
\centerline{\footnotesize Periods and Relative Rates of Period Change 
($\ell= 1$ modes) 
for selected models with $M_*= 0.50 M_{\odot}$ and 
$M_{\rm He}= 8 \times  10^{-3} M_*$}
\begin{tabular*}{185mm}{c|cc|cc|cc|cc|cc|cc|cc}
\hline
\hline
\multicolumn{1}{c|}{} & 
\multicolumn{2}{c|}{$T_{\rm eff}=28093$  K} & 
\multicolumn{2}{c|}{27029  K} &
\multicolumn{2}{c|}{26026  K} & 
\multicolumn{2}{c|}{25087  K} &
\multicolumn{2}{c|}{24025  K} & 
\multicolumn{2}{c|}{23031  K} &
\multicolumn{2}{c}{22099  K}\\
\hline
$k$ & $P$ [s] & $\dot{P}/P$ & $P$ [s] & $\dot{P}/P$ & $P$ [s] & $\dot{P}/P$ 
& $P$ [s] & $\dot{P}/P$ & $P$ [s] & $\dot{P}/P$ & $P$ [s] & $\dot{P}/P$ & $P$ [s] & $\dot{P}/P$\\
 1&        127.35&          6.90&        130.83&          5.89&        134.29&          4.77&        137.75&          3.97&        141.90&          3.18&        146.07&          2.56&        150.27&          2.06\\
 2&        164.92&          6.16&        169.04&          5.49&        173.22&          4.52&        177.50&          3.85&        182.76&          3.16&        188.13&          2.57&        193.59&          2.07\\
 3&        214.27&          5.13&        218.53&          4.18&        222.68&          3.44&        226.77&          2.85&        231.67&          2.31&        236.60&          1.88&        241.61&          1.53\\
 4&        245.71&          6.04&        251.85&          5.51&        258.27&          4.70&        264.99&          4.07&        273.38&          3.38&        282.02&          2.76&        290.83&          2.23\\
 5&        293.05&          5.98&        299.88&          4.90&        306.60&          4.04&        313.25&          3.35&        321.17&          2.68&        329.06&          2.14&        336.92&          1.70\\
 6&        320.01&          5.85&        327.78&          5.35&        336.02&          4.64&        344.67&          4.05&        355.52&          3.37&        366.75&          2.76&        378.23&          2.23\\
 7&        379.07&          5.98&        387.78&          4.81&        396.12&          3.81&        404.04&          3.04&        413.00&          2.29&        421.32&          1.71&        429.06&          1.28\\
 8&        409.50&          6.67&        420.72&          5.95&        432.38&          5.05&        444.31&          4.28&        458.69&          3.39&        472.61&          2.55&        485.12&          1.77\\
 9&        462.64&          6.41&        473.88&          5.02&        483.93&          3.55&        492.45&          2.53&        500.82&          1.67&        508.20&          1.30&        516.47&          1.29\\
10&        490.95&          5.46&        502.26&          5.12&        513.76&          4.18&        525.43&          3.53&        539.47&          2.82&        553.27&          2.21&        566.79&          1.75\\
11&        543.60&          5.05&        552.83&          3.25&        559.08&          1.81&        564.57&          1.63&        573.14&          1.88&        585.02&          2.02&        599.78&          1.91\\
12&        576.56&          3.31&        585.17&          3.52&        595.89&          3.62&        608.53&          3.43&        624.89&          2.88&        641.37&          2.27&        657.00&          1.69\\
13&        617.42&          4.19&        628.03&          3.85&        640.11&          3.67&        653.51&          3.34&        670.48&          2.77&        687.31&          2.13&        702.49&          1.49\\
14&        642.58&          4.99&        658.03&          5.55&        675.76&          4.99&        694.27&          4.25&        716.24&          3.24&        735.25&          2.03&        749.26&          1.28\\
15&        693.53&          5.80&        709.33&          4.88&        724.50&          3.75&        738.11&          2.73&        751.13&          1.68&        763.02&          1.57&        780.68&          1.93\\
16&        729.80&          6.40&        748.73&          5.54&        767.12&          4.24&        782.56&          2.77&        796.58&          1.84&        811.81&          1.84&        830.30&          1.72\\
17&        782.04&          6.04&        798.52&          4.06&        808.98&          1.88&        817.49&          1.94&        834.90&          2.71&        858.10&          2.51&        881.87&          1.94\\
18&        812.32&          5.64&        829.62&          4.30&        845.65&          3.62&        863.71&          3.47&        887.20&          2.89&        909.82&          2.10&        928.68&          1.37\\
19&        853.97&          3.16&        863.25&          2.61&        877.70&          3.69&        898.09&          3.83&        925.15&          3.19&        950.44&          2.17&        971.12&          1.56\\
20&        888.62&          2.95&        904.44&          4.55&        925.04&          4.26&        946.24&          3.48&        968.49&          2.19&        984.81&          1.44&       1006.19&          1.88\\
21&        930.53&          5.49&        952.90&          5.33&        976.07&          4.29&        996.29&          2.83&       1011.97&          1.43&       1029.22&          1.83&       1054.24&          1.89\\
22&        965.48&          6.37&        990.90&          5.68&       1015.89&          4.33&       1035.78&          2.62&       1057.17&          2.50&       1084.83&          2.38&       1113.16&          1.84\\
\hline
\end{tabular*}
NOTE -- All $\dot{P}/P$ values are in units of $10^{-16}$ s$^{-1}$.
\end{table*}

As a  variable white dwarf  cools down, its oscillation  periods ($P$)
vary in  response to evolutionary changes in  the mechanical structure
of the star.  Specifically, as the  temperature in the core of a white
dwarf decreases, the plasma increases its degree of degeneracy and the
Brunt-V\"ais\"al\"a frequency --- the most important physical quantity
in $g$-mode pulsations --- diminishes, and the pulsational spectrum of
the  star shifts  to  longer  periods.  On  the  other hand,  residual
gravitational contraction (if present) acts in the opposite direction,
thus shortening the pulsation  periods.  Competition between the
increasingly degeneracy and gravitational  contraction gives rise to a
detectable  temporal   rate  of  change   of  periods  ($\dot{P}\equiv
dP/dt$).  Roughly, the  rate  of  change of  the  pulsation period  is
related to  the rates of  change of the  temperature at the  region of
the period formation,  $\dot{T}$, and of the  stellar radius, $\dot{R_*}$,
by the expression (Winget et al. 1983):

\begin{equation}
\frac{\dot{P}}{P} \approx - a \frac{\dot{T}}{T} + 
b \frac{\dot{R_*}}{R_*},
\label{rchp}
\end{equation}

\begin{table*}
\centerline{TABLE 2}
\centerline{\footnotesize Periods and Relative Rates of Period 
Change ($\ell= 1$ modes) 
for selected models with $M_*= 0.60 M_{\odot}$ and 
$M_{\rm He}= 8 \times  10^{-3} M_*$}\begin{tabular*}{185mm}{c|cc|cc|cc|cc|cc|cc|cc}
\hline
\hline
\multicolumn{1}{c|}{} & 
\multicolumn{2}{c|}{$T_{\rm eff}=27925$  K} & 
\multicolumn{2}{c|}{26938  K} &
\multicolumn{2}{c|}{26004  K} & 
\multicolumn{2}{c|}{24950  K} &
\multicolumn{2}{c|}{23966  K} & 
\multicolumn{2}{c|}{23049  K} &
\multicolumn{2}{c}{22064  K}\\
\hline
$k$ & $P$ [s] & $\dot{P}/P$ & $P$ [s] & $\dot{P}/P$ & $P$ [s] & $\dot{P}/P$ 
& $P$ [s] & $\dot{P}/P$ & $P$ [s] & $\dot{P}/P$ & $P$ [s] & $\dot{P}/P$ & $P$ [s] & $\dot{P}/P$\\
 1&        123.16&          4.95&        126.25&          4.05&        129.36&          3.30&        133.16&          2.56&        137.01&          2.00&        140.92&          1.59&        145.61&          1.26\\
 2&        159.46&          4.93&        163.45&          4.05&        167.49&          3.30&        172.39&          2.55&        177.33&          1.98&        182.30&          1.55&        188.12&          1.19\\
 3&        197.64&          3.55&        201.20&          2.95&        204.81&          2.42&        209.21&          1.89&        213.70&          1.49&        218.20&          1.17&        223.44&          0.89\\
 4&        235.69&          5.07&        241.75&          4.17&        247.89&          3.37&        255.26&          2.58&        262.56&          1.95&        269.69&          1.47&        277.76&          1.10\\
 5&        271.80&          3.81&        277.04&          3.15&        282.33&          2.56&        288.72&          1.99&        295.23&          1.57&        301.91&          1.26&        310.08&          1.04\\
 6&        307.31&          5.00&        315.09&          4.10&        322.93&          3.30&        332.32&          2.52&        341.62&          1.91&        350.68&          1.44&        360.76&          1.04\\
 7&        347.63&          3.24&        353.24&          2.60&        358.75&          2.08&        365.30&          1.61&        371.90&          1.26&        378.60&          1.01&        386.64&          0.82\\
 8&        394.66&          4.56&        403.12&          3.28&        410.34&          2.24&        417.58&          1.44&        423.93&          1.03&        430.12&          0.82&        437.94&          0.72\\
 9&        419.51&          2.96&        426.33&          2.81&        434.42&          2.71&        445.83&          2.42&        458.55&          2.02&        471.95&          1.63&        487.89&          1.25\\
10&        460.04&          3.62&        468.28&          2.90&        476.48&          2.37&        486.64&          1.90&        497.22&          1.52&        507.85&          1.17&        519.73&          0.86\\
11&        486.33&          4.01&        496.93&          3.68&        508.45&          3.16&        522.89&          2.49&        537.17&          1.84&        550.13&          1.26&        563.16&          0.86\\
12&        533.62&          3.99&        544.07&          3.11&        553.83&          2.32&        564.13&          1.52&        572.91&          1.04&        581.80&          0.93&        595.34&          1.00\\
13&        569.46&          4.28&        581.40&          3.30&        592.27&          2.41&        604.09&          1.71&        615.98&          1.41&        628.86&          1.20&        645.01&          0.99\\
14&        608.54&          3.52&        618.12&          2.40&        627.25&          2.06&        640.25&          1.99&        655.81&          1.75&        672.03&          1.35&        689.79&          0.98\\
15&        632.97&          3.30&        644.83&          3.29&        658.72&          3.01&        676.93&          2.44&        695.17&          1.81&        711.66&          1.23&        729.36&          1.01\\
16&        673.56&          3.12&        685.30&          2.98&        698.24&          2.56&        713.68&          1.88&        727.50&          1.26&        740.44&          1.04&        759.67&          1.13\\
17&        715.79&          4.68&        732.38&          3.66&        747.30&          2.54&        761.23&          1.44&        773.06&          1.13&        787.15&          1.10&        807.43&          1.10\\
18&        752.71&          4.03&        767.22&          2.98&        780.23&          2.22&        796.58&          1.98&        815.98&          1.76&        836.16&          1.34&        858.69&          1.09\\
19&        787.73&          4.01&        801.97&          2.74&        815.41&          2.34&        834.17&          2.15&        854.89&          1.72&        874.52&          1.21&        897.69&          1.17\\
20&        818.66&          2.76&        831.58&          2.88&        848.17&          2.87&        870.67&          2.33&        892.11&          1.59&        910.63&          1.12&        935.40&          1.22\\
21&        856.30&          3.36&        872.61&          3.27&        890.45&          2.72&        910.49&          1.84&        927.66&          1.28&        946.28&          1.23&        974.56&          1.29\\
22&        900.31&          4.22&        919.46&          3.38&        936.62&          2.32&        953.00&          1.44&        969.97&          1.37&        991.07&          1.25&       1018.92&          1.22\\
23&        935.04&          4.75&        956.72&          3.60&        975.60&          2.45&        996.68&          1.98&       1020.37&          1.67&       1042.83&          1.13&       1068.49&          1.18\\
24&        970.30&          3.70&        986.29&          2.48&       1002.22&          2.41&       1027.69&          2.41&       1056.14&          1.89&       1082.72&          1.32&       1115.12&          1.33\\
\hline
\end{tabular*}
NOTE -- All $\dot{P}/P$ values are in units of $10^{-16}$ s$^{-1}$.
\end{table*}

\noindent where $a$ and $b$ are constants  whose values depend 
on  the details of  the white  dwarf modeling  ($a,b \approx  1$). The
first term in  Eq.  (\ref{rchp}) corresponds to the  rate of change in
period induced by the cooling of  the white dwarf and it is a positive
contribution, while the second term  represents the rate of change due
to gravitational  contraction and it  is a negative  contribution.  In
principle, the rate of change of period can be measured by observing a
pulsating  white dwarf  over several  seasons  when one  or more  {\it
stable} pulsation periods  are present in their light  curves.  In the
case of pulsating DA (hydrogen-rich atmospheres) white dwarfs --- also
termed  DAV or  ZZ  Ceti  variable stars  ---  cooling dominates  over
gravitational  contraction, in  such a  way  that the  second term  in
Eq. (\ref{rchp}) is negligible, and  only {\it positive} values of the
rate of  change of period  are expected.  This  is the case of  the ZZ
Ceti G117-B15A, for which Kepler et  al.  (2000) have quoted a rate of
change of the 215.2 s period of $\dot{P} = (2.3 \pm 1.4)
\times 10^{-15}$ s s$^{-1}$. In addition, Mukadam et al. (2003) report
that the  period at  213.13 s  observed in other  ZZ Ceti  star, R548,
drifts  at a  rate  $\dot{P} \leq  (5.5  \pm 1.9)  \times 10^{-15}$  s
s$^{-1}$.   For  the high  effective  temperatures characterizing  the
DOV/PNNV  instability   strip,  gravitational  contraction   is  still
significant,  to such  a degree  that its  influence on  $\dot{P}$ can
overcome the effects  of the cooling. In this case  the second term in
Eq.  (\ref{rchp})  is not  negligible  anymore  and consequently  {\it
positive} or  {\it negative} $\dot{P}$ values are  possible. For these
hot  pre-white  dwarfs,  what  determines  that  $\dot{P}$  is  either
positive or negative is the  character of the mode: if their pulsation
period corresponds  to a confined  mode\footnote{That is, a  mode with
their eigenfunctions sensitive to the  deep regions of the star.} then
a positive  value of  $\dot{P}$ is  expected, while if  the mode  is a
trapped  one\footnote{That   is,  a  mode   whose  eigenfunctions  are
concentrated mostly toward  the outer layers.} we would  expect a negative
$\dot{P}$  value (see Kawaler  \& Bradley  1994). For  PG1159-035, the
prototype star of the DOV class, a rate of $\dot{P}= (13.07 \pm 0.03)
\times 10^{-11}$ s s$^{-1}$ has been measured for the 516 s pulsation 
period,  although  uncertainties  in  the  modeling  of  its  interior
structure have led to ambiguous conclusions (Winget et al. 1985;
Costa et al.  1999).  As for  the DBV stars, in which the influence of
the gravitational contraction on $\dot{P}$ is negligible, the expected
rate  of  period  change  from theoretical  evolutionary  computations
should be {\it positive} and  of the order of $\sim 10^{-13}-10^{-14}$
s s$^{-1}$, although detections of $\dot{P}$ have not yet been made in
any DBV star.

Observational measurement  of $\dot{P}$ provides a  sensitive probe of
the structure and evolution of  white dwarf stars.  As shown by Winget
et  al. (1983),  a  measurement of  the  rate of  period  change of  a
pulsating white dwarf constitutes,  particularly for DAV and DBV white
dwarfs (which  evolve at almost constant radius)  a direct measurement
of  the cooling rate  of the  star. This,  in turn,  provides valuable
information about the core  chemical composition. Also, measurement of
the rate  of period change of  pulsating white dwarfs in  the DAV, DBV
and DOV  instability strips would allows astronomers  to calibrate the
cooling sequence  age. This, in turn,  could be employed  to infer the
age  of the galactic  disk in  the solar  neighborhood (Winget  et al.
1987).   In addition,  the measurement  of $\dot{P}$  in variable
white  dwarfs  can be  employed  to  set  constraints  on  particle
physics. McGraw et al. (1979) were  the first to  suggest that
hot pulsating white  dwarfs could be employed to  determine the effect
of neutrino cooling  as a star becomes a white  dwarf (see also Winget
et al. 1983).  The influence of neutrino energy  loss on $\dot{P}$ was
discussed in detail  by Kawaler et al. (1986) for the  case of DBV and
DOV white dwarfs. In addition,  Isern et al.  (1992) --- see also
C\'orsico  et  al. (2001a)  ---  have  explored  the effect  of  axion
emissivity  in DAV  stars. Recently,  O'Brien \&  Kawaler  (2000) have
discussed  the possibility  of inferring  limits on  the theoretically
determined  plasmon neutrino  emission  rates by  employing DOV  white
dwarfs.

\begin{table*}
\centerline{TABLE 3}
\centerline{\footnotesize Periods and Relative Rates of Period 
Change ($\ell= 1$ modes) 
for selected models with $M_*= 0.60 M_{\odot}$ and 
$M_{\rm He}= 9 \times  10^{-4} M_*$}
\begin{tabular*}{185mm}{c|cc|cc|cc|cc|cc|cc|cc}
\hline
\hline
\multicolumn{1}{c|}{} & 
\multicolumn{2}{c|}{$T_{\rm eff}= 27991$ K} & 
\multicolumn{2}{c|}{26991 K} &
\multicolumn{2}{c|}{26050 K} & 
\multicolumn{2}{c|}{24994 K} &
\multicolumn{2}{c|}{24012 K} & 
\multicolumn{2}{c|}{23095 K} &
\multicolumn{2}{c}{21967 K}\\
\hline
$k$ & $P$ [s] & $\dot{P}/P$ & $P$ [s] & $\dot{P}/P$ & $P$ [s] & $\dot{P}/P$ 
& $P$ [s] & $\dot{P}/P$ & $P$ [s] & $\dot{P}/P$ & $P$ [s] & $\dot{P}/P$ & $P$ [s] & $\dot{P}/P$\\
 1&        128.30&          6.30&        132.04&          5.16&        135.81&          4.18&        140.36&          3.22&        144.93&          2.46&        149.52&          1.90&        155.73&          1.40\\
 2&        158.94&          4.37&        162.16&          3.64&        165.43&          2.99&        169.45&          2.38&        173.58&          1.87&        177.82&          1.48&        183.70&          1.13\\
 3&        198.08&          5.46&        203.13&          4.56&        208.29&          3.75&        214.58&          2.93&        220.95&          2.26&        227.35&          1.73&        235.78&          1.23\\
 4&        237.93&          4.39&        242.76&          3.63&        247.63&          2.96&        253.49&          2.29&        259.34&          1.75&        265.14&          1.35&        272.88&          1.00\\
 5&        269.82&          5.36&        276.59&          4.50&        283.53&          3.71&        292.03&          2.91&        300.68&          2.26&        309.46&          1.76&        321.30&          1.29\\
 6&        311.14&          5.09&        318.40&          4.13&        325.56&          3.28&        333.93&          2.45&        342.04&          1.82&        349.87&          1.36&        359.91&          0.95\\
 7&        349.21&          3.97&        355.58&          3.25&        361.91&          2.63&        369.55&          2.06&        377.31&          1.61&        385.15&          1.26&        395.61&          0.92\\
 8&        391.23&          5.22&        400.40&          4.08&        408.98&          3.04&        418.16&          2.05&        426.17&          1.40&        433.64&          1.06&        444.20&          0.88\\
 9&        424.75&          3.03&        430.64&          2.52&        437.02&          2.32&        446.12&          2.19&        457.04&          1.99&        469.50&          1.70&        487.40&          1.29\\
10&        454.75&          4.40&        464.34&          3.84&        474.39&          3.22&        486.79&          2.55&        499.33&          1.96&        511.58&          1.45&        526.52&          0.93\\
11&        493.48&          4.12&        503.12&          3.54&        513.19&          2.99&        525.63&          2.36&        538.02&          1.76&        549.65&          1.26&        564.25&          0.93\\
12&        527.52&          4.81&        539.51&          4.08&        551.55&          3.25&        565.21&          2.29&        577.13&          1.51&        587.84&          1.13&        603.96&          1.04\\
13&        576.74&          4.63&        587.92&          3.21&        596.97&          2.08&        606.11&          1.49&        616.36&          1.44&        629.52&          1.40&        650.19&          1.16\\
14&        603.77&          4.09&        615.16&          3.41&        627.28&          3.02&        643.50&          2.60&        660.97&          2.09&        678.25&          1.52&        698.82&          1.02\\
15&        638.71&          3.32&        648.61&          2.88&        659.84&          2.72&        675.37&          2.38&        692.07&          1.89&        708.57&          1.42&        731.31&          1.19\\
16&        675.30&          3.89&        688.87&          3.78&        703.85&          3.24&        721.53&          2.32&        736.43&          1.42&        748.60&          0.99&        768.41&          1.12\\
17&        715.66&          5.05&        732.33&          4.10&        748.03&          3.00&        763.80&          1.86&        777.88&          1.47&        794.60&          1.40&        821.62&          1.28\\
18&        753.44&          4.67&        768.98&          3.53&        782.79&          2.52&        798.90&          2.12&        818.05&          1.95&        839.09&          1.55&        867.38&          1.25\\
19&        786.84&          4.59&        801.97&          3.21&        815.80&          2.61&        834.84&          2.43&        856.35&          1.98&        877.28&          1.41&        905.36&          1.27\\
20&        821.97&          2.86&        833.66&          2.84&        849.31&          3.05&        871.74&          2.62&        894.04&          1.86&        913.54&          1.26&        943.14&          1.34\\
21&        861.94&          3.76&        878.81&          3.69&        897.11&          3.06&        917.38&          2.02&        933.75&          1.29&        951.07&          1.27&        984.13&          1.40\\
22&        900.86&          4.99&        922.07&          4.19&        942.30&          3.05&        962.10&          1.85&        980.79&          1.61&       1002.96&          1.41&       1036.37&          1.35\\
23&        932.29&          4.99&        953.51&          3.96&        972.99&          2.88&        995.86&          2.42&       1022.65&          2.13&       1050.03&          1.55&       1085.76&          1.38\\
24&        975.02&          4.44&        992.10&          2.77&       1006.48&          2.25&       1028.72&          2.40&       1055.00&          1.95&       1080.16&          1.39&       1118.25&          1.44\\
\hline
\end{tabular*}
NOTE -- All $\dot{P}/P$ values are in units of $10^{-16}$ s$^{-1}$.
\end{table*}

More recently, D.  Winget and collaborators 
have drawn attention  the attractive possibility of employing
DBV stars as  plasmon neutrino detectors (Winget et  al.  2004). These
authors --- see also Kawaler et al.  (1986) --- have noted that in the
hotter  region of  the  DBV instability  strip,  the plasmon  neutrino
energy losses are  several times larger than the  losses due to photon
emission.  Their results suggest  that measurement of the evolutionary
period  change  in  hot  DB  white dwarf  stars  would  constitute  an
excellent probe  of the plasmon  neutrino production rates,  if period
changes in  DBV stars could  be assessed. The authors  discuss several
observational strategies  to estimate $\dot{P}$ for  DBV white dwarfs,
in addition  to the ongoing  observations of the  hot DBV EC  20058 by
Sullivan et al. (2004).

It is worth noting that the study  of Winget et al. (2004) is based on
stellar  models  for pulsating  DB  stars  with  a chemical  structure
characterized by  a pure helium envelope atop  a carbon-oxygen core,
i.e., a  single-layered envelope structure.  However, on  the basis of
evolutionary calculations  including time-dependent element diffusion,
Dehner \&  Kawaler (1995)  and Gautschy \&  Althaus (2002)  have found
that,  if DB  white dwarfs  descend from  PG1159 stars\footnote{PG1159
stars are  hydrogen-deficient post-asymptotic giant  branch (post-AGB)
stars  that are  widely  believed to  be  the result  of a  born-again
episode  (see  Sch\"onberner  1979  and  Iben et  al.  1983)  and  the
immediate  predecessors  of  most hydrogen-deficient  white  dwarfs.},
their  envelopes   would  be  characterized  by  the   presence  of  a
double-layered chemical structure.  Indeed, such  calculations show that by the
time the DB instability strip  is reached, models are characterized by
two  different  chemical  transition  zones,  i.e.,  a  double-layered
configuration.  In fact, above the carbon-oxygen core, there exists an
envelope consisting of an intershell region rich in helium, carbon and
oxygen, the  relics of the short-lived mixing  episode occurred during
the last  helium thermal pulse  that leads to the  born-again episode,
and an  overlying pure helium mantle resulting  from the gravitational
settling  of carbon and  oxygen. More  recently, Fontaine  \& Brassard
(2002) have demonstrated  that the theoretical period  spectrum of DBV
white dwarf models which  incorporate a double-layered envelope, turns
out to  be markedly distinct  from that expected for  a single-layered
configuration.   As  shown  by  these authors,  this  is  particularly
important when  attempts at constraining  the core composition  of DBs
and the  $^{12}$C($\alpha,\gamma)^{16}$O reaction rate  are made. 
This  point has  recently been addressed by  Metcalfe  et al.  (2003),
who have incorporated both the  double-layered envelope feature and adjustable
carbon-oxygen cores in DB asteroseismological fittings.

In  a recent  work, Althaus  \&  C\'orsico (2004)  have presented  new
evolutionary models of DB white  dwarf stars for various masses of the
helium   content   and  several   stellar   masses,   computed  in   a
self-consistent  way with  the predictions  of  time-dependent element
diffusion.  The initial outer layer chemical stratification assumed for 
such models corresponds to that characterizing PG1159 stars. 
By  the time  the  domain of  the  DBVs  is reached,  the
envelopes  of  these  models  are characterized  by  a  double-layered
chemical  structure  induced  by  diffusion. The  authors  find  that,
depending  on the  stellar mass,  if  DB white  dwarf progenitors  are
formed  with a  helium  content  smaller than  $\sim  10^{-3} M_*$,  a
single-layered configuration is expected to emerge {\it during} the DB
pulsation  instability strip.   As  shown in  that  paper, the  period
spacing   diagrams   exhibit   mode-trapping  substructures   when   a
double-layered configuration characterizes the envelope of the models,
substructures  that are  virtually absent  in  single-layered envelope
models.

In view  of the Winget et  al. (2004)'s claims about  the potential of
employing DBV  white dwarf stars  to place constraints on  the plasmon
neutrino  emissivity, we  judge that  the computation  of the  rate of
period change in the frame  of new stellar models which incorporate an
updated input physics, particularly time-dependent diffusion processes
as  well as  realistic  initial envelope  chemical stratifications  as
predicted by  the evolutionary history of the  progenitor stars, would
be worthwhile  to be done.  This is the  aim of the present  paper. An
additional motivation is the lack of modern tabulations of the rate of
period  change for  DBV  stars in  the  literature.  Specifically,  we
present theoretical  values for the  rates of pulsation  period change
for the effective temperature range of interest. This is done by means
of several tables providing $P$  and $\dot{P}/P\ (\equiv d \ln P /dt)$
values corresponding to dipole modes ($\ell= 1$) for different stellar
masses, effective  temperatures and thickness of  the helium envelope.
DB white dwarf evolutionary models  are the same as those presented in
Althaus  \& C\'orsico  (2004).   We compute  the  numerical values  of
$\dot{P}/P$ by  solving the equations of  linear, adiabatic, nonradial
stellar  oscillation  for evolutionary  models  representing DB  white
dwarfs.  We  briefly  examine  the  effect  of  changing  the  various
structural parameters on the theoretical rate of period change.

\begin{table*}
\centerline{TABLE 4}
\centerline{\footnotesize Periods and Relative Rates of Period Change 
($\ell= 1$ modes) 
for selected models with $M_*= 0.60 M_{\odot}$ and 
$M_{\rm He}= 1 \times  10^{-4} M_*$}
\begin{tabular*}{185mm}{c|cc|cc|cc|cc|cc|cc|cc}
\hline
\hline
\multicolumn{1}{c|}{} & 
\multicolumn{2}{c|}{$T_{\rm eff}= 27956$  K} & 
\multicolumn{2}{c|}{26901 K} &
\multicolumn{2}{c|}{25907 K} & 
\multicolumn{2}{c|}{24969 K} &
\multicolumn{2}{c|}{24086 K} & 
\multicolumn{2}{c|}{23095 K} &
\multicolumn{2}{c}{22021 K}\\
\hline
$k$ & $P$ [s] & $\dot{P}/P$ & $P$ [s] & $\dot{P}/P$ & $P$ [s] & $\dot{P}/P$ 
& $P$ [s] & $\dot{P}/P$ & $P$ [s] & $\dot{P}/P$ & $P$ [s] & $\dot{P}/P$ & $P$ [s] & $\dot{P}/P$\\
 1&        128.01&          6.60&        131.90&          5.43&        135.85&          4.42&        139.85&          3.56&        143.90&          2.84&        148.82&          2.18&        154.65&          1.62\\
 2&        165.08&          4.84&        168.74&          3.99&        172.45&          3.27&        176.21&          2.65&        180.02&          2.14&        184.64&          1.64&        190.08&          1.23\\
 3&        196.61&          4.98&        201.14&          4.18&        205.84&          3.48&        210.68&          2.87&        215.66&          2.34&        221.74&          1.80&        228.90&          1.33\\
 4&        235.43&          5.20&        241.07&          4.32&        246.82&          3.54&        252.65&          2.86&        258.49&          2.27&        265.43&          1.69&        273.34&          1.21\\
 5&        276.42&          5.28&        283.08&          4.31&        289.75&          3.47&        296.38&          2.75&        302.94&          2.17&        310.74&          1.63&        319.88&          1.22\\
 6&        310.29&          4.75&        316.96&          3.86&        323.68&          3.14&        330.44&          2.53&        337.28&          2.05&        345.66&          1.60&        355.69&          1.21\\
 7&        350.48&          4.94&        358.32&          3.98&        366.00&          3.14&        373.48&          2.44&        380.66&          1.86&        388.78&          1.32&        397.58&          0.92\\
 8&        392.79&          4.81&        401.05&          3.68&        408.79&          2.77&        416.04&          2.10&        423.03&          1.65&        431.50&          1.31&        442.12&          1.06\\
 9&        424.64&          3.58&        431.67&          3.04&        439.23&          2.70&        447.61&          2.41&        456.84&          2.10&        468.89&          1.73&        483.91&          1.35\\
10&        456.90&          4.14&        465.85&          3.61&        475.36&          3.08&        485.36&          2.58&        495.77&          2.14&        508.55&          1.65&        523.18&          1.17\\
11&        492.46&          4.54&        503.00&          3.93&        514.11&          3.31&        525.56&          2.71&        536.96&          2.11&        549.83&          1.46&        563.35&          1.01\\
12&        532.52&          4.74&        544.15&          3.93&        555.70&          3.08&        566.51&          2.26&        576.14&          1.60&        586.75&          1.19&        600.74&          1.10\\
13&        575.36&          5.07&        587.85&          3.70&        598.60&          2.51&        607.87&          1.83&        617.27&          1.61&        630.47&          1.46&        648.41&          1.26\\
14&        605.50&          4.33&        616.96&          3.35&        628.59&          2.90&        641.70&          2.65&        656.35&          2.31&        674.87&          1.80&        696.15&          1.32\\
15&        640.04&          3.14&        649.90&          2.97&        661.76&          2.89&        675.45&          2.59&        689.99&          2.13&        706.97&          1.51&        726.02&          1.22\\
16&        678.49&          4.30&        692.81&          3.95&        708.19&          3.31&        723.29&          2.49&        736.56&          1.69&        749.71&          1.10&        767.53&          1.24\\
17&        716.50&          5.11&        733.51&          4.26&        750.04&          3.20&        764.58&          2.19&        777.50&          1.67&        794.51&          1.50&        818.67&          1.45\\
18&        755.05&          5.10&        771.67&          3.75&        785.91&          2.55&        799.06&          2.11&        814.39&          2.04&        835.69&          1.71&        862.65&          1.48\\
19&        788.57&          4.54&        803.40&          3.20&        817.64&          2.75&        834.31&          2.62&        853.05&          2.25&        875.60&          1.63&        901.48&          1.42\\
20&        825.12&          3.06&        837.40&          2.94&        853.36&          3.07&        872.03&          2.72&        890.97&          2.08&        911.30&          1.37&        937.25&          1.49\\
21&        865.39&          3.91&        882.86&          3.86&        902.12&          3.23&        920.33&          2.29&        935.32&          1.49&        952.37&          1.28&        980.51&          1.56\\
22&        901.73&          5.28&        924.05&          4.45&        945.61&          3.27&        963.81&          2.14&        980.44&          1.78&       1003.48&          1.57&       1035.34&          1.58\\
23&        934.95&          5.08&        956.02&          3.91&        974.90&          2.78&        993.43&          2.45&       1015.47&          2.31&       1044.50&          1.82&       1079.78&          1.60\\
24&        977.94&          4.74&        996.39&          3.03&       1012.01&          2.40&       1030.75&          2.45&       1052.67&          2.14&       1078.27&          1.47&       1110.15&          1.59\\
\hline
\end{tabular*}
NOTE -- All $\dot{P}/P$ values are in units of $10^{-16}$ s$^{-1}$.
\end{table*}

The rest of the paper is organized  as follow. In \S 2 we describe our
evolutionary  DB white  dwarf  models, and  in  \S\ 3  we examine  the
effects of varying  the stellar mass, the mass  of the helium envelope
and the neutrino emission on the $\dot{P}/P$ values. Finally, we close
the paper with a short summary in \S 4.

\begin{figure}
\centering
\includegraphics[clip,width=250pt]{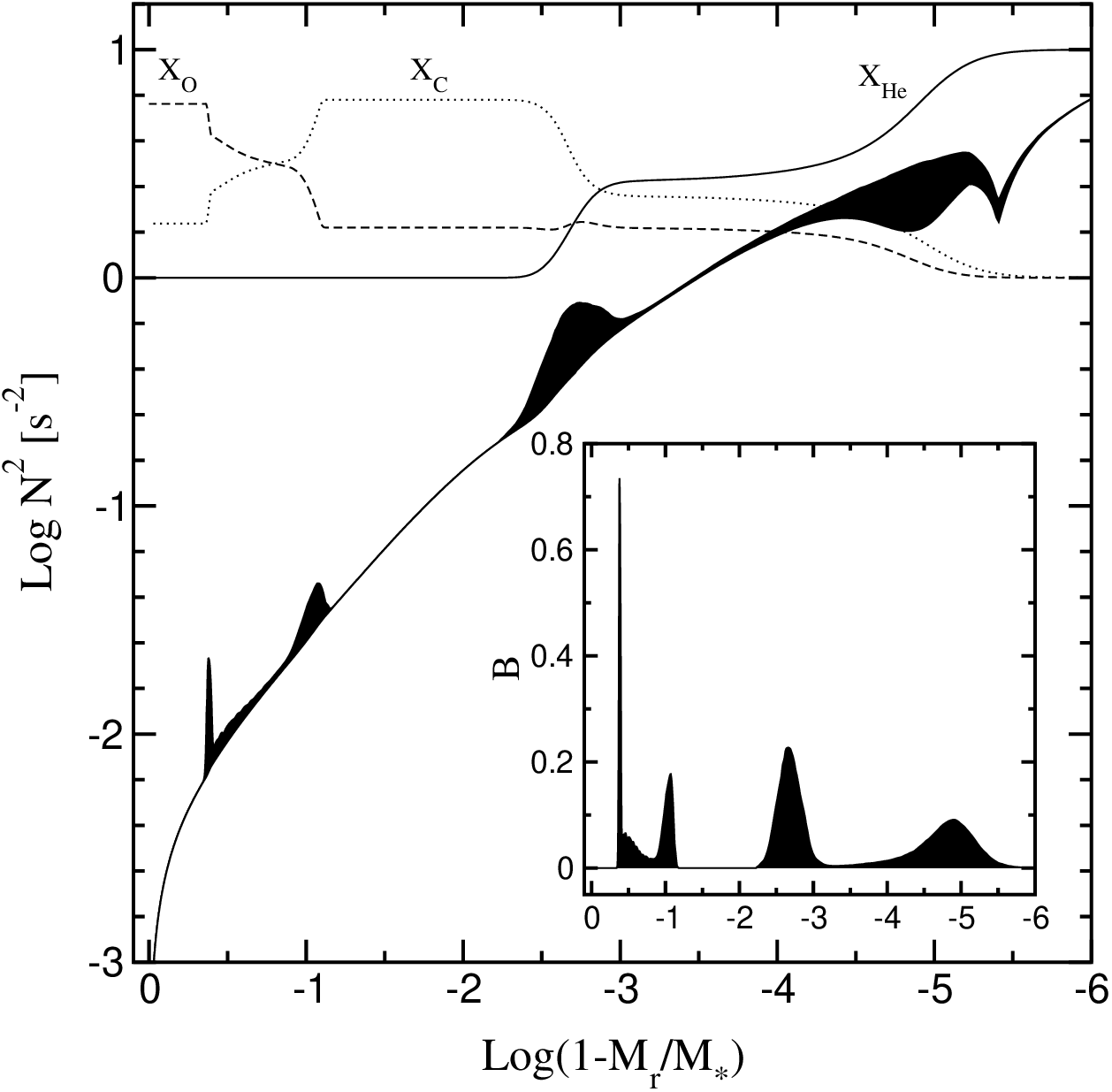}
\caption{The run of the squared Brunt-V\"ais\"al\"a frequency in terms 
of the  outer mass fraction,  corresponding to a $0.60$-\msun  DB white
dwarf model at $T_{\rm eff}= 25338$ K with a helium content of $M_{\rm
He}  \approx   9  \times  10^{-4}   M_*$.  Dark  regions   denote  the
contributions  of the  Ledoux term  $B$ (shown  in the  inset)  to the
Brunt-V\"ais\"al\"a  frequency. The chemical  profile is  displayed in
the upper zone of the plot.}
\label{fig1}
\end{figure}
  
\section{Input physics and evolutionary models}

The evolutionary models employed in  this work have been obtained with
the  DB   white  dwarf  evolutionary   code  developed  at   La  Plata
Observatory. The code is that described in Althaus \& C\'orsico (2004)
and  references  therein (see  also  Gautschy  \&  Althaus 2002).   In
particular, microphysics  includes an updated version  of the equation
of state of  Magni \& Mazzitelli (1979), OPAL  radiative opacities for
arbitrary metallicity (Iglesias \&  Rogers 1996) including carbon- and
oxygen-rich compositions,  and up-to-date neutrino  emission rates and
conductive   opacities.    In   particular,  opacities   for   various
metallicities are  required because  of the metallicity  gradient that
develops in  the envelopes as  a result of gravitational  settling. In
this work, convection is treated in the framework of the mixing length
theory  as given  by  the  ML2 parameterization  (see  Tassoul et  al.
1990).  Our evolutionary  models are calculated self-consistently with
the predictions of time-dependent element diffusion.  Our treatment of
diffusion  for multi-component  gases includes  gravitational settling
and chemical  and thermal diffusion for the  nuclear species $^{4}$He,
$^{12}$C and $^{16}$O.   As for the chemical composition  of the core,
we have adopted the chemical profiles of Salaris et al. (1997).

Our  starting stellar  configurations  correspond to  hot white  dwarf
structures  with  a  realistic  outer  layer  chemical  stratification
appropriate to that of  hydrogen-deficient PG 1159 stars. These stars,
presumed  to be  the direct  ancestors of  most DB  white  dwarfs, are
widely believed to  be the result of a  born-again episode experienced
by a  post-AGB remnant on the  early white dwarf cooling  branch. As a
result of such an episode,  most of the hydrogen content is completely
burnt,  and the  envelope is  eventually characterized  by  an uniform
chemical composition of  helium, carbon and oxygen.  In  our models we
have assumed, for such compositions,  mass fractions of 0.42, 0.36 and
0.22,  respectively, following observed  abundance patterns  in PG1159
stars.  Also, we have varied the stellar mass in the range $0.50-0.85$
\msun,  and the  mass of  the helium  content to  be $M_{\rm  He}$: $8
\times 10^{-3}, 9 \times 10^{-4}$ and $ 1 \times 10^{-4}$ $M_*$.  With
regard to the helium envelope, some words are in order.  The mass
of  the helium  envelope  is  constrained by  the  theory of  post-AGB
evolution to  be $\lesssim 10^{-2} M_*$ (Iben  1989; D'Antona \&
Mazitelli  1991).  Also, recent full  evolutionary calculations  predict the
helium content  to range from $10^{-2} M_*$  (for a 0.6  \msun\ white dwarf
remnant, see Herwig et al.  1999) to $10^{-3} M_*$ (for a 0.93
\msun\ white  dwarf, see Althaus et  al.  2003). Such  values are upper
limits in the sense that post-AGB mass loss episodes could reduce them
considerably.  In  fact, as recently emphasized by  Werner (2001), the
existence  of mass-loss rates  in the  range $10^{-7}-10^{-8}$\msun/yr
cannot be  discounted in  many luminous PG  1159 stars.   In addition,
tentative evidence for the persistence of mass-loss rates of the order
$10^{-7}-10^{-10}$
\msun/yr down to the domain of hot helium-rich white dwarfs has been
presented   (see   Werner  2001).    Thus,   the   existence  of   hot
hydrogen-deficient pre-white dwarf stars  with a helium content of the
order of $10^{-3} M_*$ could not  be discounted even in the case of DB
white dwarf remnants  with masses as low as 0.6  \msun (see Althaus \&
C\'orsico 2004).

\section{Rates of period change}

In  what  follows, we  describe  the  main  results of  our  pulsation
calculations.  We mostly concentrate on the theoretical rate of period
change.   We  assess the  $\dot{P}/P$  values  with  the help  of  the
pulsational code employed in  C\'orsico et al.  (2001b), appropriately
modified  to study  pulsating  DB white  dwarfs.   In particular,  the
treatment we follow to  assess the Brunt-V\"ais\"al\"a frequency ($N$)
is that proposed by Brassard et al.  (1991).

\begin{table*}
\centerline{TABLE 5}
\centerline{\footnotesize Periods and Relative Rates of Period Change 
($\ell= 1$ modes) 
for selected models with $M_*= 0.65 M_{\odot}$ and 
$M_{\rm He}= 9 \times  10^{-4} M_*$}
\begin{tabular*}{185mm}{c|cc|cc|cc|cc|cc|cc|cc}
\hline
\hline
\multicolumn{1}{c|}{} & 
\multicolumn{2}{c|}{$T_{\rm eff}= 28041$  K} & 
\multicolumn{2}{c|}{27090 K} &
\multicolumn{2}{c|}{26020 K} & 
\multicolumn{2}{c|}{25024 K} &
\multicolumn{2}{c|}{23943 K} & 
\multicolumn{2}{c|}{22945 K} &
\multicolumn{2}{c}{22039 K}\\
\hline
$k$ & $P$ [s] & $\dot{P}/P$ & $P$ [s] & $\dot{P}/P$ & $P$ [s] & $\dot{P}/P$ 
& $P$ [s] & $\dot{P}/P$ & $P$ [s] & $\dot{P}/P$ & $P$ [s] & $\dot{P}/P$ & $P$ [s] & $\dot{P}/P$\\
 1&        126.88&          5.11&        130.38&          4.06&        134.59&          3.05&        138.82&          2.29&        143.81&          1.69&        148.90&          1.31&        153.99&          1.05\\
 2&        153.06&          3.74&        156.19&          3.06&        160.05&          2.38&        164.01&          1.83&        168.78&          1.39&        173.71&          1.09&        178.62&          0.87\\
 3&        193.63&          4.61&        198.47&          3.69&        204.31&          2.77&        210.15&          2.08&        216.87&          1.48&        223.37&          1.08&        229.48&          0.83\\
 4&        227.51&          3.60&        231.93&          2.89&        237.25&          2.17&        242.58&          1.65&        248.81&          1.22&        255.24&          0.98&        261.98&          0.84\\
 5&        262.04&          4.58&        268.53&          3.67&        276.43&          2.78&        284.42&          2.11&        293.84&          1.56&        303.36&          1.19&        312.62&          0.92\\
 6&        300.74&          3.87&        306.90&          3.00&        314.08&          2.17&        321.07&          1.62&        329.02&          1.15&        336.68&          0.85&        343.89&          0.66\\
 7&        332.30&          3.48&        338.56&          2.81&        346.21&          2.16&        353.90&          1.62&        362.70&          1.16&        371.30&          0.88&        379.95&          0.74\\
 8&        377.94&          3.54&        384.60&          2.47&        391.56&          1.63&        398.06&          1.23&        406.14&          1.02&        415.55&          0.92&        425.99&          0.80\\
 9&        401.62&          2.99&        408.76&          2.81&        418.85&          2.48&        430.26&          2.06&        444.54&          1.58&        458.90&          1.16&        471.83&          0.82\\
10&        438.27&          3.84&        447.38&          3.09&        458.37&          2.30&        469.24&          1.71&        481.02&          1.12&        491.21&          0.75&        501.22&          0.69\\
11&        471.93&          3.67&        481.49&          3.05&        493.22&          2.31&        504.45&          1.60&        516.34&          1.08&        528.34&          0.90&        541.75&          0.84\\
12&        508.90&          4.11&        519.69&          3.02&        530.98&          1.89&        540.66&          1.32&        552.67&          1.14&        567.07&          1.01&        582.65&          0.88\\
13&        548.92&          2.53&        555.93&          1.88&        565.15&          1.75&        577.06&          1.69&        593.43&          1.36&        609.46&          0.94&        623.80&          0.76\\
14&        578.30&          3.71&        590.30&          3.15&        605.61&          2.49&        621.19&          1.85&        637.61&          1.14&        651.14&          0.77&        666.40&          0.86\\
15&        607.87&          3.18&        619.20&          2.90&        634.25&          2.36&        649.46&          1.72&        666.45&          1.24&        684.86&          1.08&        705.77&          1.01\\
16&        648.21&          4.04&        662.06&          3.09&        676.56&          1.85&        687.85&          1.16&        701.46&          1.05&        718.58&          0.96&        738.44&          0.96\\
17&        689.47&          3.71&        701.93&          2.49&        715.01&          1.79&        729.95&          1.67&        750.04&          1.30&        768.60&          0.84&        787.10&          0.92\\
18&        720.75&          3.26&        733.19&          2.61&        749.99&          2.33&        769.03&          1.89&        790.64&          1.24&        809.79&          0.90&        832.91&          1.06\\
19&        752.05&          2.99&        765.37&          2.82&        784.18&          2.43&        803.57&          1.75&        823.80&          1.15&        845.00&          1.03&        871.16&          1.09\\
20&        783.56&          3.72&        800.55&          3.31&        821.32&          2.36&        839.23&          1.47&        859.00&          1.21&        882.52&          1.05&        909.66&          1.10\\
21&        825.67&          3.79&        841.90&          2.79&        858.29&          1.68&        873.96&          1.48&        897.34&          1.35&        921.88&          0.97&        948.35&          1.08\\
22&        867.99&          3.87&        883.99&          2.49&        900.69&          1.87&        920.13&          1.68&        943.76&          1.14&        963.98&          0.79&        991.11&          1.14\\
23&        896.36&          3.53&        913.02&          2.83&        936.08&          2.56&        961.43&          1.96&        987.51&          1.15&       1011.52&          0.96&       1042.90&          1.18\\
24&        927.94&          2.72&        943.81&          2.82&        967.15&          2.42&        990.17&          1.64&       1015.52&          1.30&       1046.29&          1.18&       1081.47&          1.16\\
25&        968.77&          3.58&        988.78&          3.13&       1011.86&          2.00&       1030.38&          1.30&       1054.70&          1.24&       1082.21&          0.94&       1114.22&          1.17\\
\hline
\end{tabular*}
NOTE -- All $\dot{P}/P$ values are in units of $10^{-16}$ s$^{-1}$.
\end{table*}

\begin{figure*}
\centering
\includegraphics[clip,width=500pt]{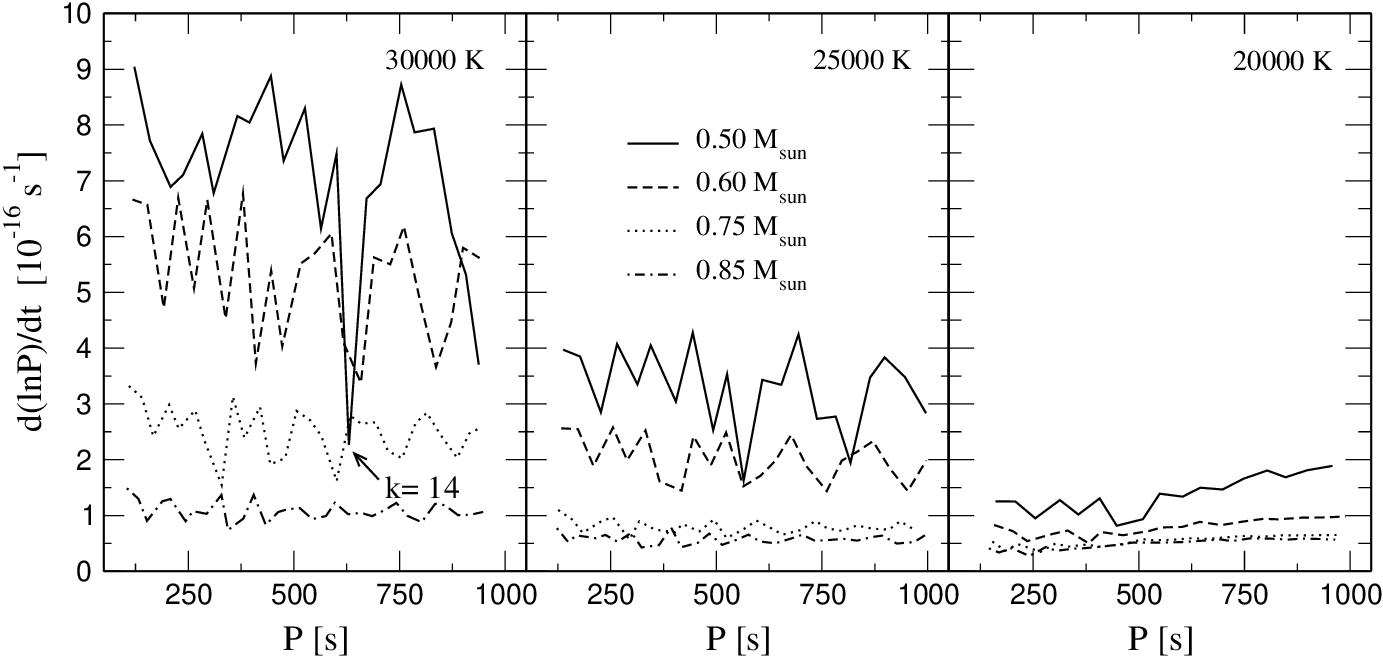}
\caption{The $\dot{P}/P$ values versus the period, 
corresponding to DB white dwarf models with stellar masses from $0.50$
to $0.85$ \msun\ (indicated with different line styles) and
a helium content  of $M_{\rm He}= 8 \times  10^{-3} M_*$. Left, centre
and  right panels display  the results  for effective  temperatures of
$30000$,  $25000$,  and $20000$  K,  respectively.   Note: the  strong
minimum  in $\dot{P}/P$  pointed  with the  arrow  is not  numerically
inflicted (see text).}
\label{fig2}
\end{figure*}

We  begin by  examining  Fig. \ref{fig1},  in  which a  representative
spatial run of  the Brunt-V\"ais\"al\"a frequency of a  DB white dwarf
is displayed.  The model is characterized by a stellar mass of $0.60$
\msun,  an effective  temperature of  $\approx 25300$  K and  a helium
content of $M_{\rm He}
\approx 9 \times 10^{-4} M_*$. In addition, the plot shows the internal 
chemical stratification of the white dwarf model, and for illustrative
purposes  the profile  of the  Ledoux term  $B$ (inset),  an important
quantity  related  to   the  computation  of  the  Brunt-V\"ais\"al\"a
frequency.  The figure emphasizes  the role of the chemical interfaces
on  the shape  of the  Brunt-V\"ais\"al\"a frequency.   In  fact, each
chemical transition region produces  clear and distinctive features in
$N$, which eventually are responsible for the mode trapping properties
of the model. At the core region, the dominant feature at $\log q
\approx  -0.4$ ($q  \equiv  1-M_r/M_*$)  is the  result  of the  steep
variation of  the oxygen/carbon  profile.  This feature  causes strong
trapping of certain modes in the core region --- `core-trapped' modes;
see Althaus  et al.  (2003) in  the context of massive  DA white dwarf
models. Such modes are characterized by an unusually large oscillation
kinetic energy. The less pronounced bump in $N$ at $\log q \approx -1$
is   much   less   relevant   to   the   structure   of   the   period
spectrum.  Finally,   in  the  envelope   of  the  model  we   find  a
double-layered  chemical   structure.  Despite  the   fact  that  this
structure is modeled by diffusion  --- and as such, characterized by a
very smooth  functional shape ---  its influence on the  mode trapping
properties  is by no  means negligible.  Indeed, Althaus  \& C\'orsico
(2004) have found that the double-layered configuration is responsible
for substructures  in the  period-spacing diagrams. 
Finally, the steep drop in the Brunt-V\"ais\"al\"a frequency at $\log q
\approx  -5.5$ is caused by the opacity change due to the  
metallicity gradient induced by diffusion in the outer layers. This 
feature occurs close  enough to the  stellar surface and has not 
appreciable effects on the model period spectrum.

In  this  work  the rate  of  period  change  is estimated  as  simple
differencing of the periods  of successive models in each evolutionary
sequence.  We  present $P$ and  $\dot{P}/P$ values corresponding  to a
modest parametric  survey of the  DB evolutionary models  presented in
Althaus \& C\'orsico (2004). In Tables 1  to 9 we provide a set of $P$
and  $\dot{P}/P$  theoretical  values  corresponding to  dipole  modes
($\ell=  1$)  for  models   with  several  stellar  masses,  effective
temperatures and thickness of the  helium envelope. We feel that these
tables could be useful for  comparison with future observations of the
rate of  period change in pulsating  DB white dwarfs. Here,  we show a
sample  of our  rate of  period change  values.  A  more comprehensive
tabulation is available at our website
\footnote{\texttt{http://www.fcaglp.unlp.edu.ar/evolgroup/}}.

Our $\ell= 1$ values of  $\dot{P}/P$ range from $\sim 4$ to $\sim
7 \times 10^{-16}$ s$^{-1}$ at  $T_{\rm eff} \approx 30000$ K and from
$\sim 1.5$ to $\sim 3 \times 10^{-16}$ s$^{-1}$ at $T_{\rm eff}
\approx 25000$ K, for periods  in the interval  $\approx$ 100--1000 s  
in the  case of a representative $0.60$-\msun\  model, irrespective of
the helium content. Our $\dot{P}/P$ values are almost twice as large as 
the values quoted by Kawaler et al. (1986) for a DB model with $0.60$
\msun\ and $M_{\rm He}\approx 3.3 \times 10^{-2} M_*$ at $T_{\rm eff} 
\approx 30000$ K. The $\ell= 1, 2$ values of  $\dot{P}$ for our models 
are  about 4  times greater  than those  of Bradley  \&  Winget (1991)
(their Table  12). This  is mainly due  to the distinct  input physics
employed  and in  particular to  the fact  that these  authors  do not
consider  neutrino   emission  in  their  Tassoul   et  al.   (1990)'s
carbon-core  DB white  dwarf models.  As compared  with the  values of
Bradley  et al. (1993) (their  Table 6,  corresponding to  a 0.6-\msun\
carbon-core white dwarf model), our  $\ell= 2$ values of $\dot{P}$ are
about  1.55 times  greater,  depending on  the effective  temperature.
Unfortunately, the  lack of extensive tabulations of  $\dot{P}$ in the
Bradley et al.  (1993)'s paper prevent us from  making a comprehensive
comparison between  their results and  ours.  We note that  our models
make use of a much more updated description of the opacities, rates of
neutrino  emission and  core chemical  composition than  that  used in
Bradley et al. (1993).

\begin{table*}
\centerline{TABLE 6}
\centerline{\footnotesize Periods and Relative Rates of Period Change 
($\ell= 1$ modes) 
for selected models with $M_*= 0.70 M_{\odot}$ and 
$M_{\rm He}= 9 \times  10^{-4} M_*$}
\begin{tabular*}{187mm}{c|cc|cc|cc|cc|cc|cc|cc}
\hline
\hline
\multicolumn{1}{c|}{} & 
\multicolumn{2}{c|}{$T_{\rm eff}= 28075$  K} & 
\multicolumn{2}{c|}{26997 K} &
\multicolumn{2}{c|}{25992 K} & 
\multicolumn{2}{c|}{25055 K} &
\multicolumn{2}{c|}{24066 K} & 
\multicolumn{2}{c|}{23033 K} &
\multicolumn{2}{c}{22077 K}\\
\hline
$k$ & $P$ [s] & $\dot{P}/P$ & $P$ [s] & $\dot{P}/P$ & $P$ [s] & $\dot{P}/P$ 
& $P$ [s] & $\dot{P}/P$ & $P$ [s] & $\dot{P}/P$ & $P$ [s] & $\dot{P}/P$ & $P$ [s] & $\dot{P}/P$\\
 1&        125.06&          3.69&        128.96&          2.73&        132.89&          2.04&        136.88&          1.59&        141.50&          1.25&        146.84&          1.00&        152.29&          0.83\\
 2&        147.72&          2.93&        151.42&          2.25&        155.22&          1.69&        159.10&          1.33&        163.60&          1.04&        168.71&          0.82&        173.82&          0.67\\
 3&        189.01&          3.36&        194.33&          2.44&        199.56&          1.77&        204.62&          1.31&        210.11&          0.97&        216.09&          0.75&        222.06&          0.63\\
 4&        217.33&          2.67&        222.22&          1.98&        227.15&          1.49&        232.17&          1.19&        238.21&          0.99&        245.69&          0.86&        253.91&          0.77\\
 5&        254.45&          3.38&        261.71&          2.49&        269.05&          1.88&        276.45&          1.45&        284.90&          1.12&        294.10&          0.82&        302.39&          0.59\\
 6&        288.43&          2.60&        294.68&          1.89&        300.90&          1.41&        306.98&          1.05&        313.58&          0.78&        320.83&          0.62&        328.75&          0.60\\
 7&        318.29&          2.81&        325.80&          2.06&        333.05&          1.46&        339.93&          1.08&        347.65&          0.85&        356.89&          0.73&        366.92&          0.64\\
 8&        359.95&          1.90&        365.57&          1.38&        371.61&          1.18&        378.66&          1.08&        388.00&          0.95&        399.26&          0.77&        410.14&          0.59\\
 9&        384.26&          3.02&        394.63&          2.44&        405.74&          1.91&        417.08&          1.46&        429.36&          1.03&        441.26&          0.67&        451.90&          0.55\\
10&        421.26&          2.73&        430.87&          1.98&        440.02&          1.37&        448.11&          0.91&        456.43&          0.70&        467.42&          0.72&        481.62&          0.74\\
11&        452.58&          2.95&        463.50&          2.06&        473.38&          1.36&        482.64&          1.05&        494.02&          0.91&        507.86&          0.74&        521.77&          0.62\\
12&        489.07&          2.35&        497.76&          1.49&        506.56&          1.28&        517.18&          1.19&        530.81&          0.99&        546.17&          0.75&        561.51&          0.67\\
13&        518.80&          2.08&        529.51&          2.01&        542.42&          1.69&        555.59&          1.24&        568.74&          0.81&        581.56&          0.60&        596.47&          0.66\\
14&        555.63&          2.93&        569.52&          2.19&        582.78&          1.48&        594.12&          0.96&        606.27&          0.80&        622.33&          0.75&        641.22&          0.75\\
15&        581.77&          2.87&        595.88&          2.11&        609.22&          1.46&        622.58&          1.21&        639.63&          1.05&        659.93&          0.83&        681.23&          0.78\\
16&        622.35&          2.44&        633.07&          1.35&        642.99&          1.14&        655.41&          1.12&        671.50&          0.91&        689.33&          0.70&        709.76&          0.77\\
17&        657.81&          2.18&        671.06&          1.94&        686.81&          1.62&        702.30&          1.12&        716.70&          0.69&        731.67&          0.60&        752.84&          0.80\\
18&        688.04&          2.70&        705.03&          2.24&        722.58&          1.62&        738.56&          1.12&        756.40&          0.93&        778.48&          0.78&        803.75&          0.84\\
19&        719.30&          2.87&        737.04&          2.14&        753.25&          1.39&        768.50&          1.12&        788.46&          1.01&        812.59&          0.81&        840.19&          0.87\\
20&        755.15&          2.95&        772.03&          1.77&        786.70&          1.31&        803.73&          1.23&        824.86&          0.95&        847.08&          0.71&        874.44&          0.88\\
21&        789.76&          2.30&        804.40&          1.70&        821.99&          1.59&        841.50&          1.23&        861.54&          0.83&        883.89&          0.75&        913.96&          0.90\\
22&        828.10&          2.13&        844.91&          1.95&        864.04&          1.49&        881.14&          0.98&        899.61&          0.83&        924.27&          0.76&        955.39&          0.90\\
23&        857.68&          2.94&        880.47&          2.37&        902.55&          1.57&        921.73&          1.11&        944.67&          0.94&        970.05&          0.69&       1001.49&          0.92\\
24&        888.14&          2.89&        909.34&          1.99&        927.99&          1.36&        948.99&          1.31&        976.68&          1.08&       1007.50&          0.83&       1044.05&          0.94\\
25&        930.34&          2.70&        948.56&          1.56&        966.30&          1.39&        987.57&          1.17&       1009.91&          0.78&       1035.44&          0.76&       1072.25&          0.94\\
26&        967.99&          2.51&        987.79&          1.90&       1010.83&          1.60&       1032.55&          1.04&       1054.27&          0.82&       1082.24&          0.73&       1119.22&          0.96\\
27&        998.51&          2.41&       1021.85&          2.23&       1047.28&          1.59&       1069.80&          1.12&       1097.37&          0.99&       1128.35&          0.72&       1167.09&          0.97\\
\hline
\end{tabular*}
NOTE -- All $\dot{P}/P$ values are in units of $10^{-16}$ s$^{-1}$.
\end{table*}

The  $\dot{P}/P$ values  in our  models exhibits  a number  of general
trends.  Firstly,  we find  that $\dot{P}/P$ generally  decreases with
decreasing effective temperature, reflecting the diminishing effect of
neutrino cooling and the increase  in the cooling timescale due to the
gradual  decreasing of  the  thermal energy  content. Second,  trapped
modes  in the helium  envelope have  smaller $\dot{P}/P$  values since
these   modes  are   concentrated   closer  to   the  surface,   where
gravitational contraction  is still appreciable.   As stated in  \S 1,
gravitational  contraction  acts  shortening  the  pulsation  periods,
causing  trapped modes  to have  smaller $\dot{P}$  values.  This
trend  is  in  agreement  with  previous studies  on  DB  white  dwarf
pulsations (see, e.g., Bradley et al. 1993).

Next,  we shall  briefly examine  the effect  of changing  the various
structural   parameters  on  the   rate  of   period  change   of  our
models. Factors that affect  $\dot{P}/P$ in white dwarf models include
the core composition, the surface chemical stratification, the stellar
mass  and  the  neutrino  emission.   Here we  restrict  ourselves  to
examining  the effects  of the  total stellar  mass, the  mass  of the
helium layer and the rate of neutrino emission.

\begin{table*}
\centerline{TABLE 7}
\centerline{\footnotesize Periods and Relative Rates of Period Change 
($\ell= 1$ modes) 
for selected models with $M_*= 0.75 M_{\odot}$ and 
$M_{\rm He}= 8 \times  10^{-3} M_*$}
\begin{tabular*}{187mm}{c|cc|cc|cc|cc|cc|cc|cc}
\hline
\hline
\multicolumn{1}{c|}{} & 
\multicolumn{2}{c|}{$T_{\rm eff}= 27969$  K} & 
\multicolumn{2}{c|}{27011 K} &
\multicolumn{2}{c|}{26002 K} & 
\multicolumn{2}{c|}{25070 K} &
\multicolumn{2}{c|}{23956 K} & 
\multicolumn{2}{c|}{23025 K} &
\multicolumn{2}{c}{22022 K}\\
\hline
$k$ & $P$ [s] & $\dot{P}/P$ & $P$ [s] & $\dot{P}/P$ & $P$ [s] & $\dot{P}/P$ 
& $P$ [s] & $\dot{P}/P$ & $P$ [s] & $\dot{P}/P$ & $P$ [s] & $\dot{P}/P$ & $P$ [s] & $\dot{P}/P$\\
 1&        115.07&          2.03&        118.26&          1.61&        121.98&          1.30&        125.83&          1.10&        130.99&          0.92&        135.77&          0.81&        141.33&          0.71\\
 2&        147.18&          1.88&        150.91&          1.46&        155.16&          1.14&        159.35&          0.92&        164.61&          0.72&        169.11&          0.59&        173.95&          0.49\\
 3&        173.02&          1.43&        176.28&          1.08&        179.89&          0.83&        183.44&          0.69&        188.17&          0.60&        192.78&          0.57&        198.63&          0.55\\
 4&        213.57&          1.72&        218.49&          1.33&        224.12&          1.06&        229.91&          0.90&        237.66&          0.76&        244.68&          0.65&        252.41&          0.53\\
 5&        236.16&          1.64&        241.58&          1.36&        248.16&          1.14&        255.04&          0.97&        264.13&          0.79&        272.26&          0.67&        281.49&          0.59\\
 6&        277.38&          1.66&        283.33&          1.19&        289.30&          0.80&        294.32&          0.57&        300.19&          0.45&        305.85&          0.45&        313.36&          0.46\\
 7&        297.53&          1.40&        303.14&          1.10&        309.88&          0.95&        317.46&          0.89&        328.60&          0.82&        339.35&          0.73&        351.84&          0.63\\
 8&        336.46&          1.03&        341.60&          0.95&        348.47&          0.87&        355.90&          0.74&        365.24&          0.56&        372.78&          0.44&        380.97&          0.40\\
 9&        372.43&          2.01&        382.53&          1.55&        393.38&          1.09&        402.32&          0.70&        411.34&          0.48&        419.17&          0.44&        429.07&          0.45\\
10&        396.26&          1.41&        403.35&          0.99&        410.92&          0.79&        419.88&          0.85&        435.03&          0.87&        450.46&          0.79&        468.69&          0.70\\
11&        435.31&          1.27&        442.23&          0.91&        450.48&          0.81&        459.61&          0.71&        471.35&          0.55&        481.21&          0.46&        493.41&          0.51\\
12&        457.41&          1.47&        467.69&          1.40&        480.84&          1.17&        494.06&          0.93&        509.82&          0.68&        522.69&          0.55&        538.12&          0.58\\
13&        495.52&          1.50&        505.36&          1.11&        515.23&          0.75&        524.02&          0.60&        536.72&          0.60&        550.82&          0.62&        570.27&          0.68\\
14&        528.02&          1.73&        539.38&          1.17&        550.72&          0.86&        562.24&          0.74&        577.67&          0.61&        591.10&          0.51&        608.65&          0.63\\
15&        556.80&          1.27&        566.46&          1.06&        579.67&          1.03&        594.65&          0.91&        613.88&          0.69&        629.75&          0.56&        650.51&          0.69\\
16&        583.05&          1.41&        595.15&          1.26&        609.90&          1.02&        624.25&          0.79&        642.26&          0.66&        660.27&          0.66&        685.76&          0.76\\
17&        619.08&          1.47&        631.67&          1.16&        644.78&          0.80&        656.76&          0.66&        673.88&          0.62&        690.99&          0.58&        715.23&          0.74\\
18&        659.60&          1.62&        672.23&          1.00&        684.72&          0.78&        698.55&          0.73&        717.55&          0.60&        733.91&          0.51&        759.14&          0.79\\
19&        683.63&          1.48&        697.58&          1.24&        715.78&          1.12&        734.86&          0.90&        757.29&          0.64&        776.21&          0.57&        804.98&          0.81\\
20&        715.69&          1.25&        729.03&          1.16&        746.06&          0.96&        762.71&          0.77&        785.46&          0.72&        808.91&          0.69&        841.61&          0.82\\
21&        746.97&          1.59&        763.68&          1.28&        781.17&          0.88&        797.14&          0.72&        819.76&          0.66&        841.48&          0.60&        874.08&          0.84\\
22&        782.59&          1.56&        797.99&          1.07&        814.10&          0.86&        832.45&          0.82&        857.35&          0.65&        878.77&          0.56&        912.26&          0.85\\
23&        818.75&          1.44&        833.23&          1.01&        851.15&          0.93&        870.18&          0.76&        893.50&          0.60&        915.85&          0.60&        951.68&          0.85\\
24&        846.51&          1.55&        865.32&          1.35&        888.22&          1.06&        908.75&          0.75&        933.41&          0.62&        956.87&          0.57&        994.35&          0.89\\
25&        873.67&          1.35&        891.63&          1.25&        913.03&          0.97&        935.22&          0.90&        968.32&          0.80&        997.63&          0.65&       1038.55&          0.90\\
26&        915.25&          1.52&        933.00&          1.04&        950.10&          0.76&        969.13&          0.74&        995.97&          0.63&       1022.32&          0.65&       1065.09&          0.86\\
27&        952.28&          1.64&        971.17&          1.09&        992.43&          0.95&       1014.28&          0.73&       1039.24&          0.55&       1063.09&          0.55&       1105.62&          0.92\\
28&        980.15&          1.47&       1000.22&          1.27&       1026.48&          1.08&       1051.18&          0.80&       1082.11&          0.67&       1109.78&          0.57&       1154.29&          0.93\\
\hline
\end{tabular*}
NOTE -- All $\dot{P}/P$ values are in units of $10^{-16}$ s$^{-1}$.
\end{table*}

\begin{figure*}
\centering
\includegraphics[clip,width=500pt]{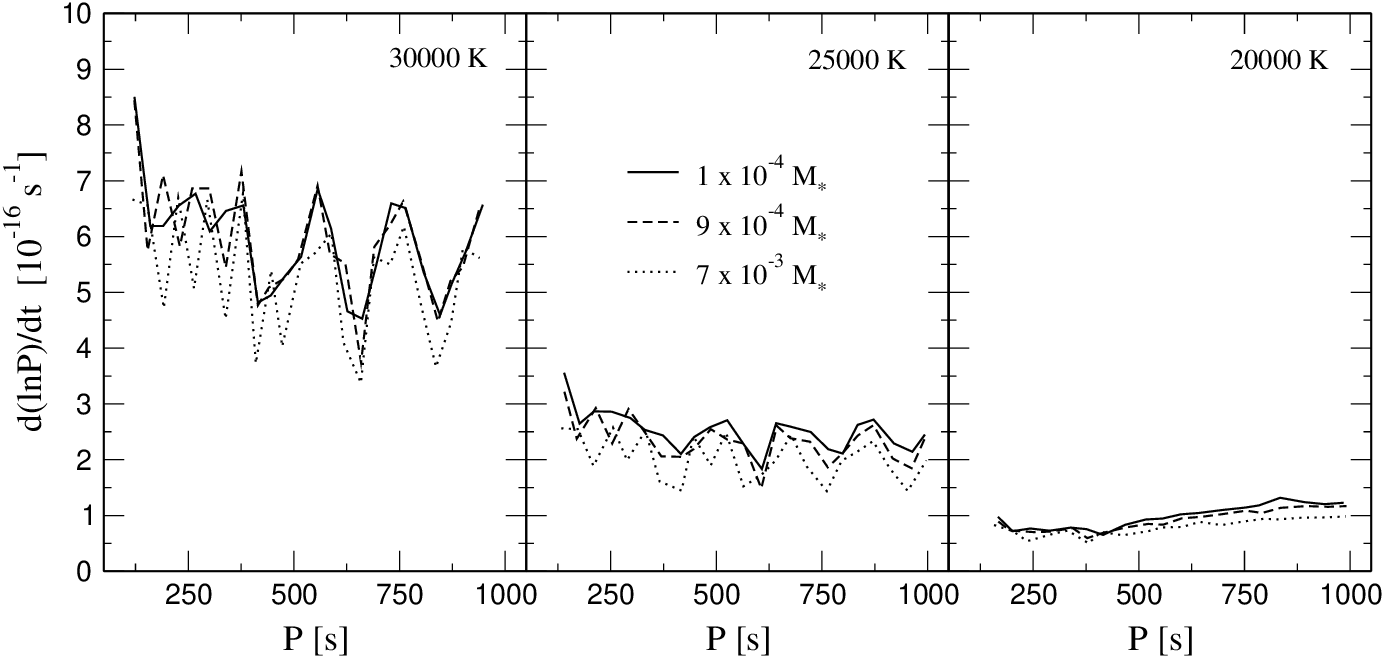}
\caption{The $\dot{P}/P$ values versus the period, corresponding to 
DB  white dwarf  models with  a stellar  mass of  $0.60$ \msun\  and a
content of helium of $1  \times 10^{-4} M_*$ (solid lines),
$9  \times 10^{-4}  M_*$ (dashed  lines), and  $8 \times
10^{-3} M_*$ (dotted  lines). Left, centre and right panels
display the  results for  effective temperatures of  $30000$, $25000$,
and  $20000$  K,  respectively.  $\dot{P}/P$ values  corresponding  to
models  with thick  helium envelopes  are only  marginally  smaller as
compared with the case of thin helium envelopes.}
\label{fig3}
\end{figure*}

\noindent{\sl Effect of the stellar mass:}\
$\dot{P}/P$ in white dwarf  stars is very sensitive mostly to the
stellar mass.  The effect of  the stellar mass  on the rate  of period
change has been explored by Kawaler et al. (1986) for DB white dwarfs,
by Kawaler et al. (1985a) for DO white dwarfs, and by Bradley \& Winget
(1991),  Bradley  et  al.  (1993)  and Bradley  (1996)  for  DA  white
dwarfs. Lower  mass models  exhibit larger $\dot{P}/P$  values through
the range of  effective temperature of pulsating DB  white dwarf stars
(Kawaler et al. 1986). This can be understood on the basis that, if we
consider a  fixed $T_{\rm  eff}$ value, lower  mass white  dwarfs have
larger  luminosities and  lower  total heat  capacity,  and thus  cool
faster with a larger $\dot{P}$.   This can be also understood in terms
of the simple, yet accurate  enough for our purposes, cooling model of
Mestel (1952). Within the framework  of the Mestel (1952) cooling law,
Kawaler et al. (1986) have  derived a relation between $\dot{P}/P$ and
the stellar mass:

\begin{equation}   
\frac{d \ln P}{d t}= 2 \times 10^{-30} A \left( \frac{\mu}{\mu_e^2} 
\right)^{0.286}\ \left(\frac{M}{M_{\odot}} \right)^{-1.190}\ T_{\rm eff}^
{2.857},
\label{eq2}
\end{equation}

\noindent where $\mu$ is the mean molecular weight, $\mu_e$ the molecular 
weight  per  electron, and  $A$  the atomic  mass  of  the ions.   In
particular,  it  is  convenient   to  note  that  from  our  numerical
computations the magnitude of the  period derivative varies by about a
factor of $6-7$  in the stellar mass interval  $0.50-0.85 M_\odot$, as
we show in  Fig. \ref{fig2} for DB models with a  helium layer mass of
$\approx 8 \times 10^{-3} M_*$.   Each panel in the figure illustrates
the  situation  at  different  effective  temperatures  covering  the
observed DBV  instability strip.  As we  can see from  the figure, the
general theoretical expectations from Eq. (\ref{eq2}) are borne out by
our  detailed numerical  computations.  In fact,  the  rate of  period
change  for the  less massive  models is  always larger  than  for the
more massive ones.

\begin{figure}
\centering
\includegraphics[clip,width=260pt]{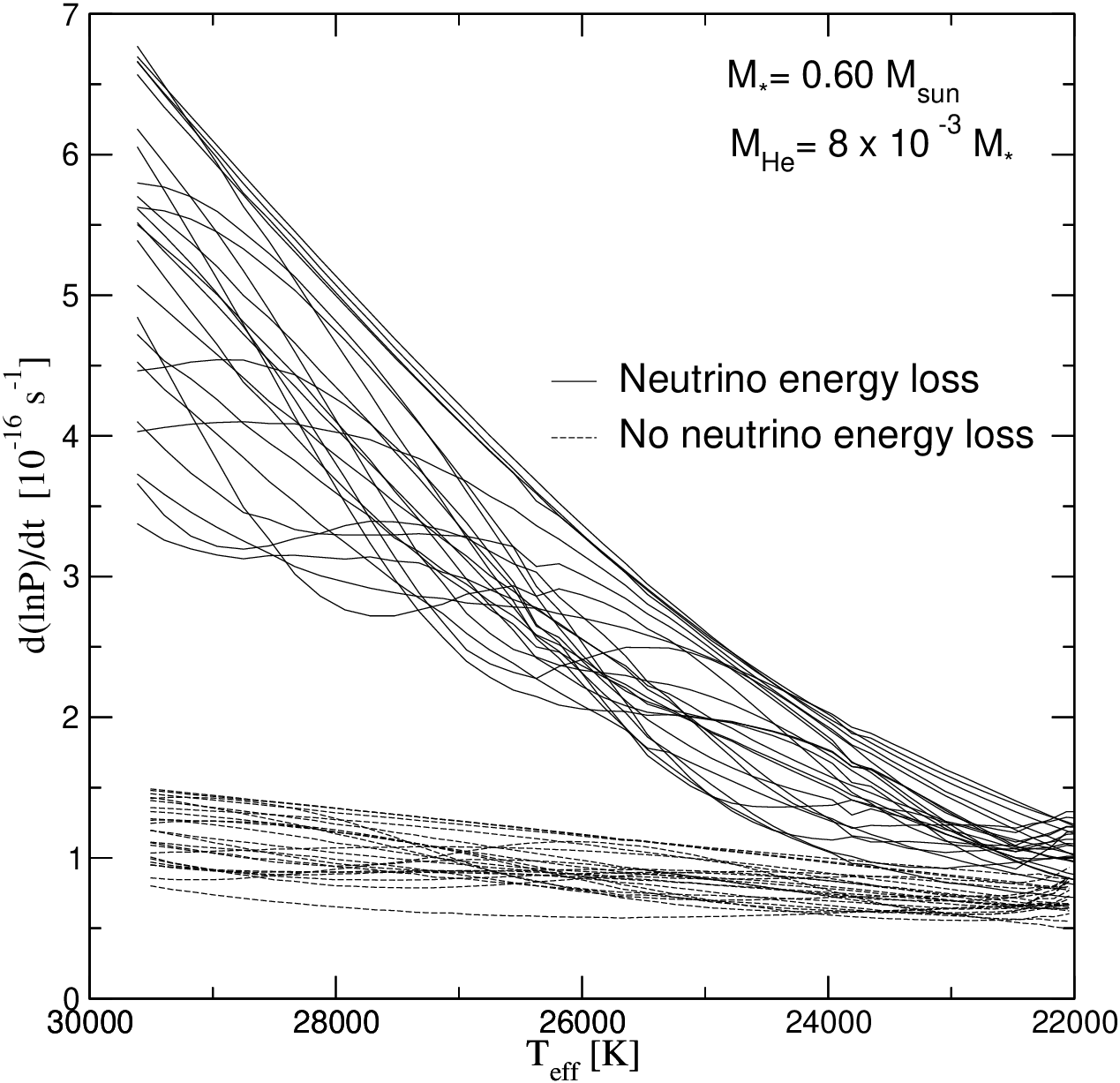}
\caption{The evolution of the relative rate of period change including 
(continuous lines)
and ignoring (short-dashed lines) the  effect of neutrino, in terms of
the effective temperature, corresponding  to a sequence of 0.60-\msun\
white dwarf models ($M_{\rm He}= 8 \times 10^{-3} M_*$).}
\label{fig4}
\end{figure}

Fig.    \ref{fig2}   further    emphasizes   the   complex   structure
characterizing  the $\dot{P}/P$-distribution:  no  matter the  stellar
mass is, the diagrams show  abundant local maxima and minima. The mode
with $k= 14$  corresponding to the $0.50$-\msun\ white  dwarf model at
$T_{\rm eff} \approx 30000$ K (pointed with the arrow in left panel of
the figure)  is striking.  The very pronounced  minima in the  rate of
period change of this mode is not the result of a numerical glitch; we
have  verified the  reality  of this  by  examining their  oscillation
kinetic energy and eigenfunctions, and we have found that this mode is
strongly  trapped in  the  helium-rich envelope.  As  we have  already
anticipated, trapped modes like this one must have a small $\dot{P}/P$
value,  reflecting  primarily the  contraction  of  the outer  layers,
rather  than carrying information  about the  processes linked  to the
cooling  of the  white  dwarf   (Bradley  et al.  1993). We  have
verified that almost all the trapped modes in the helium-rich envelope
of our models correspond to local minima in $\dot{P}/P$.

\noindent{\sl Effect of the helium layer mass:}\
Now we  examine the possible  dependence of the $\dot{P}/P$  values on
the thickness of the helium layer, holding the stellar mass fixed.  We
have considered  the following values  for the helium mass:  $1 \times
10^{-4} M_*$, $9 \times 10^{-4}  M_*$, and $8 \times 10^{-3} M_*$.  It
is worth recalling that in the sequence of DB models with $M_{\rm He}
\approx  1 \times  10^{-4} M_*$  the envelope  is characterized  by an
initially   double-layered   configuration    which   evolves   to   a
single-layered structure before the model  reaches the red edge of the
DBV  instability strip (Althaus  \& C\'orsico  2004).  The  effects of
varying the mass of the helium layer, keeping the total mass constant,
are  illustrated in  Fig.   \ref{fig3}.  This  figure  shows that  the
$\dot{P}/P$  values  corresponding   to  models  with  massive  helium
envelopes are  slightly smaller  than the case  of thinner  ones. This
effect  can be explained  on the  basis that  our models  with massive
helium  layers  cool slightly  more  slowly  as  compared with  models
characterized by  thinner helium envelopes. This is  understood on the
basis  that models  with massive  (and also  more  transparent) helium
envelope  are characterized  by  lower central  temperatures which  in
turn, implies  that such models  have initially an excess  of internal
energy to  get rid  of --- see  Tassoul et  al. (1990). This  leads to
longer  cooling age  at this  stage and  thus smaller  rate  of period
variations.

\begin{table*}
\centerline{TABLE 8}
\centerline{\footnotesize Periods and Relative Rates of Period Change 
($\ell= 1$ modes) 
for selected models with $M_*= 0.75 M_{\odot}$ and 
$M_{\rm He}= 9 \times  10^{-4} M_*$}
\begin{tabular*}{178mm}{c|cc|cc|cc|cc|cc|cc|cc}
\hline
\hline
\multicolumn{1}{c|}{} & 
\multicolumn{2}{c|}{$T_{\rm eff}= 28041$  K} & 
\multicolumn{2}{c|}{27039 K} &
\multicolumn{2}{c|}{25990 K} & 
\multicolumn{2}{c|}{25029 K} &
\multicolumn{2}{c|}{24001 K} & 
\multicolumn{2}{c|}{23033 K} &
\multicolumn{2}{c}{21997 K}\\
\hline
$k$ & $P$ [s] & $\dot{P}/P$ & $P$ [s] & $\dot{P}/P$ & $P$ [s] & $\dot{P}/P$ 
& $P$ [s] & $\dot{P}/P$ & $P$ [s] & $\dot{P}/P$ & $P$ [s] & $\dot{P}/P$ & $P$ [s] & $\dot{P}/P$\\
 1&        123.03&          2.40&        126.67&          1.81&        130.85&          1.40&        135.08&          1.15&        140.02&          0.94&        145.06&          0.79&        150.74&          0.65\\
 2&        142.97&          2.02&        146.53&          1.53&        150.59&          1.18&        154.64&          0.95&        159.26&          0.76&        163.89&          0.64&        169.15&          0.55\\
 3&        183.88&          2.04&        188.41&          1.48&        193.25&          1.07&        197.86&          0.84&        203.16&          0.70&        208.78&          0.63&        215.67&          0.58\\
 4&        207.81&          1.77&        212.37&          1.37&        217.80&          1.12&        223.67&          0.99&        231.05&          0.87&        239.05&          0.77&        248.49&          0.66\\
 5&        246.86&          2.18&        253.53&          1.66&        261.12&          1.26&        268.46&          0.97&        276.19&          0.69&        282.82&          0.49&        289.35&          0.38\\
 6&        275.46&          1.67&        281.04&          1.23&        287.05&          0.89&        292.70&          0.70&        299.52&          0.64&        307.70&          0.65&        318.76&          0.64\\
 7&        306.39&          1.78&        312.76&          1.24&        319.61&          0.94&        326.72&          0.82&        335.56&          0.72&        344.83&          0.61&        355.10&          0.50\\
 8&        340.44&          1.33&        346.67&          1.23&        355.13&          1.10&        364.26&          0.91&        374.49&          0.70&        383.97&          0.54&        394.41&          0.48\\
 9&        371.00&          2.22&        381.23&          1.69&        392.44&          1.20&        402.31&          0.83&        412.13&          0.61&        422.03&          0.55&        434.67&          0.55\\
10&        403.20&          1.61&        410.54&          1.04&        417.71&          0.74&        425.52&          0.74&        437.16&          0.77&        450.95&          0.71&        467.73&          0.65\\
11&        434.69&          1.68&        443.23&          1.20&        453.24&          1.00&        463.96&          0.85&        476.07&          0.65&        487.31&          0.50&        500.83&          0.55\\
12&        464.85&          1.40&        474.03&          1.33&        486.22&          1.12&        498.52&          0.88&        511.98&          0.68&        525.31&          0.58&        542.31&          0.63\\
13&        496.77&          2.00&        508.98&          1.48&        521.24&          0.94&        531.50&          0.68&        543.46&          0.62&        557.39&          0.60&        576.65&          0.68\\
14&        533.33&          1.79&        544.05&          1.15&        555.10&          0.90&        567.52&          0.84&        583.26&          0.72&        598.64&          0.56&        618.17&          0.69\\
15&        557.91&          1.73&        569.60&          1.32&        584.14&          1.14&        599.77&          0.95&        617.57&          0.75&        635.09&          0.63&        658.86&          0.77\\
16&        590.63&          1.24&        601.20&          1.22&        615.44&          1.03&        629.50&          0.79&        645.04&          0.65&        662.22&          0.63&        687.30&          0.76\\
17&        629.79&          1.92&        644.36&          1.36&        658.04&          0.81&        669.64&          0.64&        684.65&          0.62&        701.79&          0.57&        727.55&          0.81\\
18&        660.68&          1.96&        675.90&          1.36&        692.33&          1.07&        710.02&          0.93&        729.99&          0.68&        747.86&          0.53&        775.34&          0.86\\
19&        690.22&          1.68&        703.43&          1.18&        720.07&          1.08&        738.48&          0.91&        759.95&          0.76&        782.97&          0.69&        815.68&          0.85\\
20&        722.13&          1.45&        736.73&          1.36&        755.60&          1.09&        773.17&          0.78&        792.29&          0.66&        813.65&          0.63&        846.69&          0.87\\
21&        754.41&          1.81&        772.02&          1.44&        790.43&          0.95&        807.32&          0.79&        829.33&          0.73&        852.75&          0.63&        887.27&          0.89\\
22&        791.81&          1.82&        808.43&          1.20&        825.20&          0.92&        844.17&          0.85&        866.73&          0.67&        888.69&          0.58&        924.45&          0.90\\
23&        825.01&          1.98&        843.38&          1.31&        864.24&          1.09&        885.33&          0.83&        907.13&          0.62&        930.08&          0.59&        968.71&          0.94\\
24&        850.86&          1.48&        867.65&          1.35&        890.98&          1.18&        914.75&          0.93&        942.67&          0.80&        971.92&          0.68&       1013.92&          0.93\\
\hline
\end{tabular*}
NOTE -- All $\dot{P}/P$ values are in units of $10^{-16}$ s$^{-1}$.
\end{table*}

\begin{figure*}
\centering
\includegraphics[clip,width=500pt]{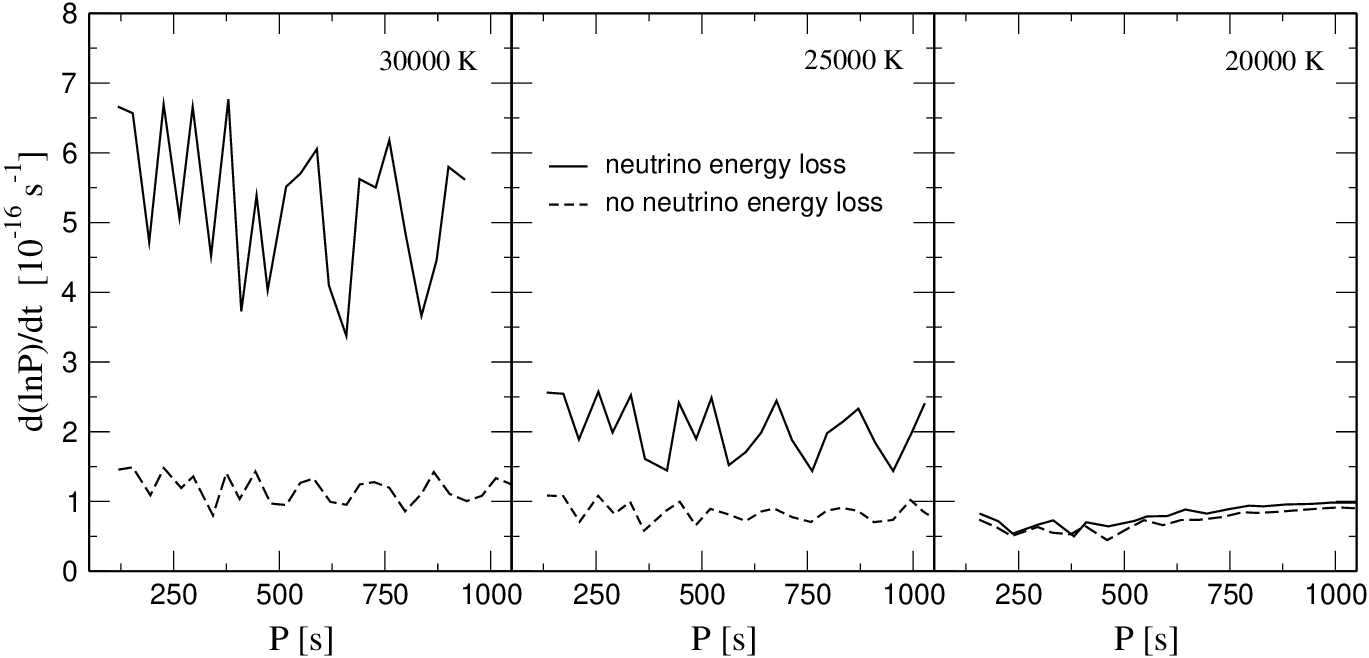}
\caption{The relative rate of period change in terms of the period 
corresponding to the same models 
as in  Fig. \ref{fig4}, at effective temperatures  of $\approx 30000$,
$\approx 25000$ and $\approx 20000$ K.  Continuous (dashed) 
lines correspond to computations  including (ignoring) the effect of
neutrinos.}
\label{fig5}
\end{figure*}

\noindent {\sl Effect of the neutrino emission:}\
Finally, we  shall examine  the effect of  the neutrino emission on the
$\dot{P}/P$ values. We begin by examining Fig. \ref{fig4}, in which we
show  the evolution  of the  rate  of period  change in  terms of  the
effective temperature corresponding to a sequence of 0.60-\msun\ white
dwarf models  with a helium content  of $M_{\rm He}=  8 \times 10^{-3}
M_*$.  Continuous  lines  correspond to  the  case in  which
neutrino emission has been included in our DB models, and short-dashed
lines correspond to the situation  in which neutrino emission has been
ignored\footnote{It  is  worth  mentioning  that we  have  turned  off
neutrino  emission at  a high  effective temperature  ($\approx 60000$
K).  As a  result, the  model  has adequately  readjusted its  thermal
structure long before the DBV regime is reached.}. At the hot edge of
the  DB  instability  strip,  the values  of  $\dot{P}/P$  considering
neutrino emission are greater by a  factor of about 5 as compared with
calculations  in  which neutrino  emission  has  not  been taken  into
account.  At $T_{\rm  eff} \approx  25000$ K  the factor  is  of about
2.5. In Fig.  \ref{fig5} we illustrate $\dot{P}/P$  versus periods at
three effective  temperatures, for the  case in which  neutrino losses
have been considered (continuous lines) and for the situation in which
neutrino  emission  has  been  ignored (dashed  lines).  Clearly,  the
$\dot{P}/P$   values  are   very  sensitive   to   neutrino  emission,
particularly  at the  high effective  temperatures  characterizing the
blue edge of the DBV  instability strip.  Thus, we essentially recover
the  results  reported by  Winget  et  al.  (2004).  We  conclude,  in
agreement with Winget  et al. (2004)'s claims, that  at high effective
temperatures within the DB instability strip the eventual detection of
the  rate  of  period change  in  DBV  white  dwarfs could  allow  the
astronomers  to constrain  the  production rates  of plasmon  neutrino
emission.

\begin{table*}
\centerline{TABLE 9}
\centerline{\footnotesize Periods and Relative Rates of Period Change 
($\ell= 1$ modes) 
for selected models with $M_*= 0.85 M_{\odot}$ and 
$M_{\rm He}= 8 \times  10^{-3} M_*$}
\begin{tabular*}{186mm}{c|cc|cc|cc|cc|cc|cc|cc}
\hline
\hline
\multicolumn{1}{c|}{} & 
\multicolumn{2}{c|}{$T_{\rm eff}= 27968$ K} & 
\multicolumn{2}{c|}{27040 K} &
\multicolumn{2}{c|}{26010 K} & 
\multicolumn{2}{c|}{25003 K} &
\multicolumn{2}{c|}{24014 K} & 
\multicolumn{2}{c|}{23038 K} &
\multicolumn{2}{c}{21960 K}\\
\hline
$k$ & $P$ [s] & $\dot{P}/P$ & $P$ [s] & $\dot{P}/P$ & $P$ [s] & $\dot{P}/P$ 
& $P$ [s] & $\dot{P}/P$ & $P$ [s] & $\dot{P}/P$ & $P$ [s] & $\dot{P}/P$ & $P$ [s] & $\dot{P}/P$\\
 1&        110.97&          1.11&        114.43&          0.99&        118.52&          0.87&        122.70&          0.77&        126.95&          0.68&        131.23&          0.60&        136.07&          0.53\\
 2&        138.36&          0.89&        141.68&          0.74&        145.34&          0.61&        148.88&          0.53&        152.39&          0.47&        155.97&          0.43&        160.18&          0.40\\
 3&        157.44&          0.69&        160.61&          0.66&        164.70&          0.65&        169.33&          0.64&        174.42&          0.61&        179.87&          0.57&        186.28&          0.52\\
 4&        197.77&          0.97&        203.09&          0.84&        209.16&          0.71&        215.00&          0.59&        220.48&          0.49&        225.60&          0.41&        231.07&          0.34\\
 5&        218.20&          0.96&        223.92&          0.82&        230.50&          0.72&        237.27&          0.65&        244.43&          0.61&        252.06&          0.58&        261.21&          0.53\\
 6&        250.83&          0.53&        254.69&          0.51&        259.78&          0.52&        265.70&          0.53&        272.30&          0.51&        279.35&          0.48&        287.60&          0.43\\
 7&        272.95&          0.97&        280.58&          0.89&        289.71&          0.79&        298.95&          0.69&        308.06&          0.59&        316.93&          0.51&        326.78&          0.45\\
 8&        305.06&          0.72&        310.74&          0.56&        316.74&          0.46&        322.75&          0.43&        329.28&          0.42&        336.65&          0.43&        346.15&          0.43\\
 9&        343.00&          0.72&        349.43&          0.58&        356.75&          0.52&        364.30&          0.47&        371.99&          0.42&        379.78&          0.38&        389.45&          0.40\\
10&        356.75&          0.90&        366.63&          0.92&        379.39&          0.87&        392.95&          0.78&        406.73&          0.68&        419.94&          0.56&        434.29&          0.50\\
11&        394.17&          0.67&        401.17&          0.55&        408.87&          0.46&        416.67&          0.43&        425.18&          0.43&        435.31&          0.48&        450.50&          0.56\\
12&        425.26&          0.91&        435.47&          0.73&        446.45&          0.60&        457.01&          0.51&        467.35&          0.45&        477.68&          0.41&        492.31&          0.51\\
13&        446.89&          0.67&        456.16&          0.71&        468.92&          0.72&        483.14&          0.68&        497.84&          0.60&        512.53&          0.53&        531.22&          0.57\\
14&        481.73&          0.78&        491.86&          0.65&        502.81&          0.53&        513.53&          0.47&        524.89&          0.46&        537.71&          0.48&        557.25&          0.59\\
15&        507.76&          0.90&        520.29&          0.77&        534.50&          0.66&        548.43&          0.56&        562.03&          0.49&        575.61&          0.45&        596.25&          0.59\\
16&        534.22&          0.72&        545.23&          0.68&        559.60&          0.68&        575.86&          0.66&        593.21&          0.60&        610.84&          0.54&        635.62&          0.65\\
17&        563.02&          0.75&        575.07&          0.69&        589.20&          0.60&        603.37&          0.53&        618.17&          0.50&        634.77&          0.53&        660.20&          0.63\\
18&        599.01&          0.78&        611.48&          0.64&        625.34&          0.55&        639.44&          0.50&        654.11&          0.47&        669.82&          0.47&        696.61&          0.67\\
19&        625.97&          0.92&        641.62&          0.78&        659.54&          0.68&        677.42&          0.59&        694.88&          0.50&        712.38&          0.48&        742.25&          0.70\\
20&        651.11&          0.71&        665.00&          0.72&        683.43&          0.70&        703.55&          0.66&        724.56&          0.60&        746.35&          0.56&        778.67&          0.69\\
21&        684.35&          0.78&        699.15&          0.68&        716.09&          0.59&        733.41&          0.54&        752.23&          0.53&        773.39&          0.55&        807.38&          0.71\\
22&        714.41&          0.84&        730.59&          0.70&        748.78&          0.61&        767.61&          0.56&        787.18&          0.51&        807.93&          0.52&        843.11&          0.71\\
23&        744.68&          0.75&        760.36&          0.68&        780.00&          0.65&        800.68&          0.58&        821.42&          0.51&        843.04&          0.51&        879.78&          0.71\\
24&        775.02&          0.82&        792.77&          0.73&        813.23&          0.62&        833.46&          0.55&        854.48&          0.51&        877.27&          0.52&        916.02&          0.72\\
25&        799.37&          0.88&        819.63&          0.81&        843.63&          0.71&        867.58&          0.61&        890.98&          0.53&        914.71&          0.52&        956.48&          0.75\\
26&        831.80&          0.73&        848.10&          0.62&        868.29&          0.62&        892.12&          0.64&        919.11&          0.62&        948.18&          0.60&        991.63&          0.71\\
27&        869.63&          0.71&        886.65&          0.63&        906.96&          0.56&        927.36&          0.50&        948.68&          0.47&        973.21&          0.53&       1017.88&          0.75\\
28&        896.10&          0.93&        919.68&          0.82&        945.10&          0.63&        968.17&          0.52&        990.78&          0.47&       1014.94&          0.49&       1062.17&          0.77\\
29&        922.59&          0.80&        943.71&          0.74&        970.29&          0.72&        999.84&          0.67&       1028.90&          0.55&       1056.64&          0.52&       1105.71&          0.76\\
30&        955.46&          0.80&        975.90&          0.67&       1000.25&          0.63&       1026.53&          0.59&       1055.53&          0.59&       1088.05&          0.59&       1139.19&          0.74\\
31&        986.08&          0.75&       1007.39&          0.71&       1033.53&          0.63&       1060.00&          0.56&       1087.32&          0.52&       1117.51&          0.56&       1170.46&          0.76\\
\hline
\end{tabular*}
NOTE -- All $\dot{P}/P$ values are in units of $10^{-16}$ s$^{-1}$.
\end{table*}

\section{Summary}

The rate of period change  is the most exciting observable quantity in
pulsating white  dwarf stars because it potentially  provides a direct
measure of the rate of cooling of white dwarfs, giving astronomers the
possibility  of inferring the  age of  the galactic  disk in  the solar
neighborhood (Winget  et al. 1987). Also, measurement  of $\dot{P}$ in
white  dwarfs  offers  an  unique  opportunity  to  place  interesting
constraints on  neutrino physics (O'Brien  \& Kawaler 2000;  Winget et
al.  2004)  and  to test  any  additional  sink  of energy  (Isern  et
al. 1992; C\'orsico  et al. 2001a).  In addition,  the small magnitude
of $\dot{P}$ characterizing pulsating white dwarfs implicates that its
detection  would also  impose strong  constraints on  the  presence of
other mechanisms acting on shorter time scales and that can modify the
pulsation periods, such as stellar  rotation (Kawaler et al. 1985b) and
binary orbital motion (Kepler et al. 1991).

Rates of  period change have been  detected for two members  of the ZZ
Ceti  variable class (G117-B15A  and R548)  and only  for a single DOV white
dwarf (PG 1159-035).  In the case of ZZ  Ceti stars, the observed
values for  the rate of period  change are in good  agreement with the
theoretical expectations.  As for  PG 1159-035, however,  the observed
value is largely in excess  as compared to the theoretical predictions
(see Costa  et al. 1999).  For DBV stars  no detection has  yet been
made,  but this  situation could  soon  be changed by the  ongoing
observations of the  hot DBV EC 20058 by Sullivan et  al. (2004). In a
recent paper,  Winget et al.   (2004) have called  attention to
the potential of employing DBV  white dwarf stars to place constraints
on  the  plasmon  neutrino  emissivity.  Prompted  by   Winget  et
al.  (2004)'s claims,  we have  computed in  this work  the  change of
period  rate in  pulsating  DB  white dwarf  stars  employing the  new
stellar   models   recently   presented   in  Althaus   \&   C\'orsico
(2004). These  stellar models incorporate an updated  input physics in
addition to a self-consistent  treatment of the time-dependent element
diffusion to assess the shape  of the chemical interfaces --- an issue
of fundamental importance in the context of white dwarf pulsations. In
view of the lack of modern  tabulations of $\dot{P}/P$ in DBV stars in
the literature, we have  included tables providing $P$ and $\dot{P}/P$
values corresponding to dipole modes ($\ell= 1$) for different stellar
masses,   effective   temperatures  and   thickness   of  the   helium
envelope. We  have also examined  the effects of varying  the neutrino
emission rate on  the $\dot{P}/P$ values. We find,  in agreement with
previous studies, that the rate  of period change is a function mainly
of the  total stellar mass --- there  is more than a  factor 6 between
the $\dot{P}/P$ values for  0.50- and 0.85-\msun\ models. At variance,
the rates of period change  are rather insensitive to the thickness of
the helium  envelope.  We also  find that, within the  DBV instability
strip,  $\dot{P}/P$  is  strongly  dependent on  the  neutrino  energy
losses. Specifically, the  $\dot{P}/P$ values are considerably smaller
(by a factor  of about $3-5$) when neutrino losses  are ignored in our
models.  Our  results are  consistent  with  the conclusion  arrived  at
recently   by  Winget  et   al.  (2004)   about  the   feasibility  of
quantitatively constraining  the production rate of  plasmon neutrinos by
employing pulsating DB white dwarfs.

\begin{acknowledgements}

We wish to acknowledge the  suggestions and comments of an anonymous referee
that strongly improved  the original version of this  work.  We warmly
acknowledge  E.  Garc\'{\i}a-Berro  for   a  careful  reading  of  the
manuscript.  This   research  was   supported  by  the   Instituto  de
Astrof\'{\i}sica La Plata. L.G.A.  acknowledges the Spanish MCYT for a
Ram\'on y Cajal Fellowship.

\end{acknowledgements}

\end{document}